\documentclass[10pt,aps,prd,reprint,nofootinbib,superscriptaddress]{revtex4-1}

\usepackage{amsmath}
\usepackage{graphicx}
\usepackage[colorlinks,pdfusetitle,linktocpage=true]{hyperref}
\hypersetup{colorlinks=true,allcolors=[rgb]{1,0.56,0}}

\newcommand{\orcid}[1]{\href{https://orcid.org/#1}{#1}}
\newcommand{\e}[1]{\times10^{#1}}

\begin{document}

\title{Individual Neutrino Masses From a Supernova}

\author{Peter B.~Denton}
\email{pdenton@bnl.gov}
\thanks{\orcid{0000-0002-5209-872X}}
\affiliation{High Energy Theory Group, Physics Department, Brookhaven National Laboratory, Upton, NY 11973, USA}

\author{Yves Kini}
\email{y.kini@uva.nl}
\thanks{\orcid{0000-0002-0428-8430}}
\affiliation{Anton Pannekoek Institute for Astronomy, University of Amsterdam, Science Park 904, 1090GE Amsterdam, the Netherlands}

\begin{abstract}
A nearby supernova will carry an unprecedented wealth of information about astrophysics, nuclear physics, and particle physics.
Because supernova are fundamentally neutrino driven phenomenon, our knowledge about neutrinos -- particles that remain quite elusive -- will increase dramatically with such a detection.
One of the biggest open questions in particle physics is related to the masses of neutrinos.
Here we show how a galactic supernova provides information about the masses of each of the three mass eigenstates \emph{individually}, at some precision, and is well probed at JUNO.
This information comes from several effects including time delay and the MSW effect within the supernova.
The time delay feature is strongest during a sharp change in the flux such as the neutronization burst; additional information may also come from a QCD phase transition in the supernova or if the supernova forms a black hole.
We consider both standard cases as dictated by local oscillation experiments as well as new physics motivated scenarios where neutrino masses may differ across the galaxy.
\end{abstract}

\date{May 2, 2025}

\maketitle

\tableofcontents

\section{Introduction}
Understanding the nature of neutrinos is one of the main open questions in particle physics.
Since the discovery that neutrinos oscillate and have mass \cite{Super-Kamiokande:1998kpq,SNO:2001kpb,SNO:2002tuh}, at least seven new parameters have been added to our model of particle physics: three masses and four parameters governing the mixing between the mass and flavor (interaction) eigenstates.
Of these seven, six are probable in oscillations. The one remaining parameter is effectively the absolute neutrino mass scale.
Considerable progress has been made in measuring these six oscillation parameters and an approximate picture has emerged; current and future measurements are expected to complete the measurements of these parameters \cite{Denton:2022een,deGouvea:2022gut}.

Accessing the final remaining parameter, the absolute neutrino mass scale, can proceed in various ways.
The most competitive and likely fastest way to get at it is via cosmological measurements of the early Universe combined with galaxy surveys \cite{Loverde:2024nfi}, while additional orthogonal measurements come from tritium end point measurements \cite{Katrin:2024tvg}.

Another method of probing the absolute neutrino mass scale is via measurements of neutrinos from a galactic supernova (SN) \cite{Zatsepin:1968kt}, see also \cite{Loredo:2001rx,Nardi:2003pr,Nardi:2004zg,Pagliaroli:2010ik,Lu:2014zma,Hyper-Kamiokande:2018ofw,Hansen:2019giq,Pompa:2022cxc,Pitik:2022jjh,Brdar:2022vfr,Parker:2023cos}, which is the focus of this paper.
Existing studies typically assume equal masses for each neutrino, incompatible with oscillation results.
We develop the theory here, for the first time, for the impact of three separate mass states on both the propagation of neutrinos from a SN to the Earth, and also the impact within the SN.
We consider the JUNO detector \cite{JUNO:2023dnp}, under construction now, as the target due to its large volume and low thresholds.
Finally, we consider three possible sharp features with which to observe the time-delay signature and thus extract the masses of neutrinos: the neutronization burst \cite{Mirizzi:2015eza} which is quite likely to exist, a QCD phase transition inside the SN \cite{Sagert:2008ka,Fischer:2010wp,Fischer:2017lag,Zha:2021fbi,Fischer:2021tvv,Kuroda:2021eiv,Bauswein:2022vtq,Lin:2022lck,Pitik:2022jjh} which may exist, and finally the possibility that the SN forms a black hole (BH) \cite{Sekiguchi:2010ja,Gullin:2021hfv} which happens for some subset of SN explosions.

\begin{figure}
\centering
\includegraphics[width=\columnwidth,trim=4 33 35 35,clip]{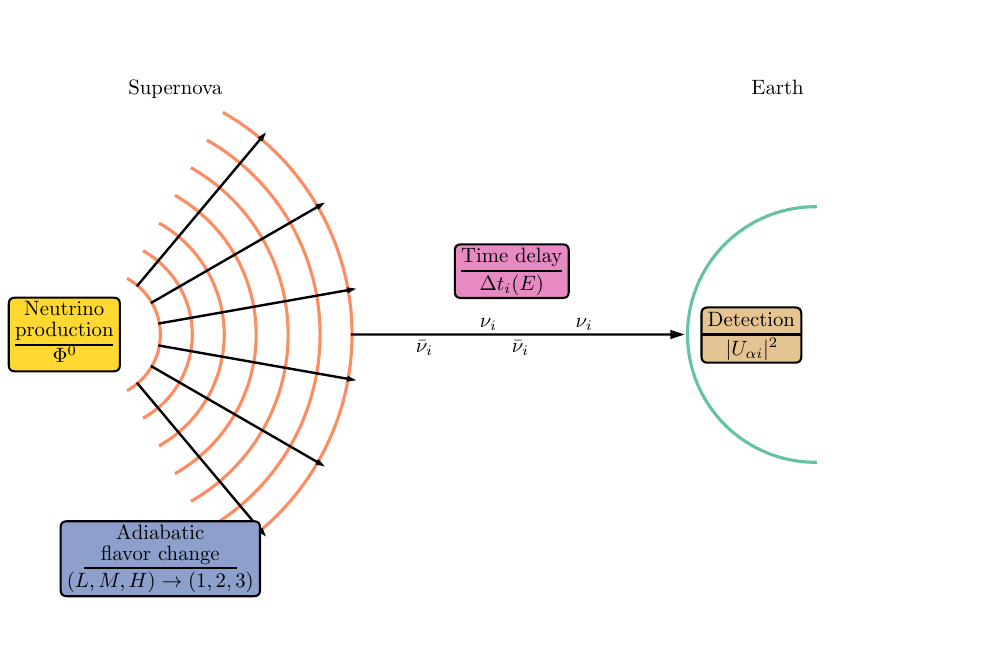}
\caption{A schematic showing the effects in play from the SN explosion, propagation, and detection.
Different masses will potentially affect the adiabatic flavor change in the SN; they will affect the propagation as a function of neutrino energy, and then each mass stage is projected back to a given flavor state at the Earth.}
\label{fig:propagation schematic}
\end{figure}

In this article, we first discuss the possibility that neutrinos may have different masses in the region of space where a SN is likely to go off in section \ref{sec:spatial neutrino masses} to motivate a range of benchmarks.
We then review the physics of neutrinos from a SN including for the first time the impact of each individual mass eigenstate, and a careful check of the role of jump probabilities in section \ref{sec:sn neutrinos}.
In section \ref{sec:time delay} we discuss the features in the SN spectrum that will lead to a noticeable time delay signature.
A schematic of the total process is shown in fig.~\ref{fig:propagation schematic}.
Section \ref{sec:detection} discusses the experimental details considered and our numerical results are presented in section \ref{sec:sensitivities}.
Finally we discuss our results in section \ref{sec:discussion} and conclude in section \ref{sec:conclusions}.

\section{Possibility of Spatially Evolving Neutrino Masses}
\label{sec:spatial neutrino masses}
While the smallest possible neutrino masses from oscillations and the largest neutrino masses allowed by cosmology are all somewhat small in the scheme of the discussion presented here and are expected to be at the $\sim0.01-0.1$ eV level \cite{Denton:2023hkx}, we also consider the possibility that neutrino mass generation has a spatial dependence due to perhaps a coupling to dark matter (DM).
In the presence of such an interaction, neutrinos would have different, presumably larger\footnote{We note that it is also possible that neutrinos are \textit{lighter} near the galactic center due to an interaction with DM if neutrinos have a bare mass term with a different sign from the DM interaction and a cancellation occurs.}, masses near the galactic center where the DM density is about two to three orders of magnitude larger than near our solar system \cite{Linden:2014sra}.
Near the galactic center is where a SN is most likely to occur and thus the masses there may not even be consistent with the $\Delta m^2$'s measured in neutrino oscillation experiments on the Earth and in our solar system.
We also note that cosmological constraint on the sum of neutrino masses comes primarily from early periods of the Universe and when the neutrinos are primarily in the vacuum of space in the redshift range $10\lesssim z\lesssim100$, not in the centers of DM halos \cite{Lorenz:2021alz}.

Various versions of such models exist and they have been discussed in a variety of different contexts.
We will briefly review several, but as the specific details do not matter for our results, we will not go into detail.

In \cite{Davoudiasl:2018hjw} they proposed that neutrino masses are exactly zero except near DM due to a long-range mediator.
Thus the measured parameters would be related to the local DM density and the neutrino masses would be higher in the galactic center proportional to the DM density there and the individual masses would remain proportional to those near the Earth, but the neutrino mass separations would become much larger.
We note here that the additional presence of a bare mass term arising from any typical scenario (e.g.~seesaw) with possibly a different sign, could lead to a wide range of nontrivial phenomenology, masses, and mass differences.
For example, it could lead to mass orderings completely different from the two standard ones: the normal ordering $m_1<m_2<m_3$ and the inverted ordering $m_3<m_1<m_2$.
In fact, any of the six possible mass orderings can be realized in these new physics scenarios in the galactic center and still be consistent with oscillation data on the Earth.

In \cite{Ge:2019tdi,Choi:2019zxy,Choi:2020ydp,Sen:2023uga,SevillanoMunoz:2024ayh} various additional such models in different contexts were also proposed.
These models also show the opportunities for different scaling laws between DM density and neutrino masses.
Thus considering a wide range of neutrino masses at the point of the SN is justified as reasonably possible, up to the limit from SN1987A \cite{Kamiokande-II:1987idp,Bionta:1987qt,Alekseev:1987ej} which is $m_\nu\lesssim6-9$ eV \cite{Loredo:2001rx}.
Moreover, since the Large Magellanic Cloud, the host of the SN, is in an environment with less DM than the galactic center, the neutrino masses inferred from a galactic SN could in principle be even higher than the limit from SN1987A.

Ref.~\cite{Ge:2024ftz} also considers a generic model modifying neutrino masses and points out that if neutrino masses depend on local DM density, then the effective time delay in a galactic SN must be integrated depending on the geometry of the location of the SN within the galaxy.
We do not consider the impact of varying masses during propagation as that can be well approximated by the appropriate average effective mass during propagation.
A realistic experimental analysis of time-delay with neutrinos on the heavier side, however, would need to consider this effect.

\section{Neutrinos From a Supernova}
\label{sec:sn neutrinos}
Supernovae are the most efficient neutrino factories in the Universe where $\sim10^{58}$ neutrinos are produced over a short period of time with energies in the tens of MeV.
This makes SN prime environments to study many aspects of physics that are otherwise challenging to probe, including neutrino properties.

\subsection{Neutrino Mixing}
It is well established \cite{Dighe:1999bi} that the flux of neutrinos at the Earth of each neutrino flavor in the typical scenario is given by
\begin{widetext}
\begin{align}
\Phi_{\nu_e}(E,t)={}&|U_{e3}|^2\Phi_{\nu_e}^0(E,t)+(|U_{e1}|^2+|U_{e2}|^2)\Phi_{\nu_x}^0(E,t) \label{eq:Phinue NO}\\
\Phi_{\bar\nu_e}(E,t)={}&|U_{e1}|^2\Phi_{\bar\nu_e}^0(E,t)+(|U_{e2}|^2+|U_{e3}|^2)\Phi_{\bar\nu_x}^0(E,t)\\
2\Phi_{\nu_x}(E,t)={}&(|U_{\mu3}|^2+|U_{\tau3}|^2)\Phi_{\nu_e}^0(E,t)+(|U_{\mu1}|^2+|U_{\tau1}|^2+|U_{\mu2}|^2+|U_{\tau2}|^2)\Phi_{\nu_x}^0(E,t)\\
2\Phi_{\bar\nu_x}(E,t)={}&(|U_{\mu1}|^2+|U_{\tau1}|^2)\Phi_{\bar\nu_e}^0(E,t)+(|U_{\mu2}|^2+|U_{\tau2}|^2+|U_{\mu3}|^2+|U_{\tau3}|^2)\Phi_{\bar\nu_x}^0(E,t)
\end{align}
for the normal mass ordering and 
\begin{align}
\Phi_{\nu_e}(E,t)={}&|U_{e2}|^2\Phi_{\nu_e}^0(E,t)+(|U_{e1}|^2+|U_{e3}|^2)\Phi_{\nu_x}^0(E,t)\\
\Phi_{\bar\nu_e}(E,t)={}&|U_{e3}|^2\Phi_{\bar\nu_e}^0(E,t)+(|U_{e1}|^2+|U_{e2}|^2)\Phi_{\bar\nu_x}^0(E,t)\\
2\Phi_{\nu_x}(E,t)={}&(|U_{\mu2}|^2+|U_{\tau2}|^2)\Phi_{\nu_e}^0(E,t)+(|U_{\mu1}|^2+|U_{\tau1}|^2+|U_{\mu3}|^2+|U_{\tau3}|^2)\Phi_{\nu_x}^0(E,t)\\
2\Phi_{\bar\nu_x}(E,t)={}&(|U_{\mu3}|^2+|U_{\tau3}|^2)\Phi_{\bar\nu_e}^0(E,t)+(|U_{\mu1}|^2+|U_{\tau1}|^2+|U_{\mu2}|^2+|U_{\tau2}|^2)\Phi_{\bar\nu_x}^0(E,t) \label{eq:Phinuebar IO}
\end{align}
\end{widetext}
for the inverted mass ordering and we have assumed the standard Mikheyev-Smirnov-Wolfenstein (MSW) \cite{Wolfenstein:1977ue,Mikheyev:1985zog} behavior within the SN.
Here $\Phi_{\nu_\alpha}(E,t)$ is the flux at the Earth of neutrino flavor $\nu_\alpha$ with energy $E$ and at time $t$.
The superscript $\Phi^0$ indicates that it is the flux at the source, but after applying the appropriate scaling factors for propagation to the Earth.
The flavor states $\nu_x$ and $\nu_y$ (not shown above) are the non-$\nu_e$ states that exist in the basis that is modified from the usual flavor basis via the $U_{23}(\theta_{23})$ rotation.

This structure is written out in this fashion that may seem somewhat obtuse, but the usage of e.g.~$|U_{e1}|^2+|U_{e2}|^2$ instead of just $1-|U_{e3}|^2$ which follows by unitarity, is necessary to allow one to calculate the impact of time delay for each mass state separately.

Before we handle the completely arbitrary mass scenario, it is important to define the PMNS \cite{Pontecorvo:1957cp,Maki:1962mu} matrix carefully.
While there are multiple possible definitions of the mass states, see \cite{Denton:2020exu,Denton:2021vtf} for a discussion of this, we use the most self-consistent definition
\begin{equation}
|U_{e1}|^2>|U_{e2}|^2>|U_{e3}|^2\,.
\label{eq:normalcy}
\end{equation}
This definition is quite useful due to the precise measurement of electron neutrino disappearance from Daya Bay, RENO, and KamLAND \cite{DayaBay:2022orm,RENO:2018dro,KamLAND:2013rgu}.
At this stage, we make the ansatz that while the three neutrino masses may vary arbitrarily, the mixing matrix remains unchanged.

We make a new definition of the three mass states as heavy, medium, and light:
\begin{equation}
m_H>m_M>m_L\,.
\end{equation}
This allows us to create a single framework that covers not only both the normal mass ordering ($L=1$, $M=2$, and $H=3$) and the inverted mass ordering ($L=3$, $M=1$, $H=2$), but also any other possible configuration.

We assume (and will soon show) that neutrinos are produced with an energy and in a density larger than all resonances for any available $\Delta m^2$.
Thus the neutrinos will evolve through both level crossings.
Under this set of reasonable and justified assumptions, in general, we find
\begin{widetext}
\begin{align}
\Phi_{\nu_e}(E,t)={}&|U_{eH}|^2\Phi_{\nu_e}^0(E,t)+(|U_{eL}|^2+|U_{eM}|^2)\Phi_{\nu_x}^0(E,t)\label{eq:Phinue arbitrary}\\
\Phi_{\bar\nu_e}(E,t)={}&|U_{eL}|^2\Phi_{\bar\nu_e}^0(E,t)+(|U_{eM}|^2+|U_{eH}|^2)\Phi_{\bar\nu_x}^0(E,t)\label{eq:Phinuebar arbitrary}\\
2\Phi_{\nu_x}(E,t)={}&(|U_{\mu H}|^2+|U_{\tau H}|^2)\Phi_{\nu_e}^0(E,t)+(|U_{\mu L}|^2+|U_{\tau L}|^2+|U_{\mu M}|^2+|U_{\tau M}|^2)\Phi_{\nu_x}^0(E,t)\label{eq:Phinux arbitrary}\\
2\Phi_{\bar\nu_x}(E,t)={}&(|U_{\mu L}|^2+|U_{\tau L}|^2)\Phi_{\bar\nu_e}^0(E,t)+(|U_{\mu M}|^2+|U_{\tau M}|^2+|U_{\mu H}|^2+|U_{\tau H}|^2)\Phi_{\bar\nu_x}^0(E,t)\label{eq:Phinuxbar arbitrary}
\end{align}
\end{widetext}
While these expressions (eqs.~\ref{eq:Phinue arbitrary}-\ref{eq:Phinuxbar arbitrary}) look similar to those above (eqs.~\ref{eq:Phinue NO}-\ref{eq:Phinuebar IO}), the careful placement of the mass orderings allows for a completely general description of neutrino masses and is necessary for a time delay calculation involving each neutrino mass state separately both for assumed true values of the neutrino masses, as well as alternate test values when computing statistical significances. These equations enable the calculation of the neutrino flux at Earth for any mass ordering and represent an important result of this work.

We now confirm that the necessary resonance conditions are met.
From \cite{Dighe:1999bi}, we have that the resonance density is on the order of
\begin{multline}
\rho_{\rm res}\simeq1.4\e6{\rm\ g/cc}\\\times\left(\frac{\Delta m^2}{1{\rm\ eV}^2}\right)\left(\frac{10{\rm\ MeV}}E\right)\left(\frac{0.5}{Y_e}\right)\cos2\theta\,.
\end{multline}
The largest this can be, for our problem, is for $m_i\sim1$ eV and $E\sim1$ MeV, which leads to a maximum resonance density of $\sim10^7$ g/cc.
Since the neutrinos decouple at densities of $\sim10^{11}-10^{12}$ g/cc, neutrinos are always safely produced well above these resonances, and the above formulas are correct.
In fact, these assumptions are still valid for $m_i$ up to $\sim10$ eV or greater.

There is one possible addition to the above expressions which are due to jump probabilities.
Jump probabilities happen near resonances where the eigenvalues are close to each other in a rapidly evolving environment, see \cite{Mikheyev:1985zog,Parke:1986jy}.

The final check that is required is to confirm that jump probabilities do not dramatically alter this picture.

\subsection{Jump Probabilities}
Jump probabilities typically appear all over the expressions in the previous subsection connecting different mass states with each other; the above expressions assume all these jump probabilities are zero for any combinations of masses.
We prove this here.

\subsubsection{Basic Picture of Jump Probabilities}
For small angles we have
\begin{equation}
P_j=\exp\left(-\frac\pi2\gamma\right)\,,
\label{eq:Pjump simple}
\end{equation}
at the resonance, where the adiabaticity parameter\footnote{We typically assume that $n_e\propto\rho$, although it is possible that $Y_e$ depends on $r$; we ignore this.} $\gamma>0$ is
\begin{equation}
\gamma=\frac{\Delta m^2}{2E}\frac{s_2^2}{c_2}\frac1{\dot n_e/n_e}\,.
\end{equation}

We also note that the mixing angles present in the PMNS matrix are not particularly small.
For this and other reasons, a more complete expression for the jump probabilities is
\begin{equation}
P_j=\frac{\exp\left(-\frac\pi2\gamma f\right)-\exp\left(-\frac\pi2\gamma\frac f{s^2}\right)}{1-\exp\left(-\frac\pi2\gamma\frac f{s^2}\right)}\,,
\end{equation}
where $s$ is $\sin\theta$ for the relevant mixing angle in the two-flavor picture and $f$ is some new function.
In the limit where the mixing angle is small, eq.~\ref{eq:Pjump simple} is recovered.
This equation is still somewhat approximate, but the remaining corrections are not likely to be relevant for a SN.

Our calculation here follows that from \cite{Kuo:1988pn}.
We suppose that the density profile of a SN follows the functional form
\begin{equation}
\rho=Ar^n\,,
\end{equation}
for some scaling factor $A$ and some power $n$ (note that the sign of $n$ varies in the literature).
Then the $f$ term is given by
\begin{equation}
f=2\sum_{m=0}^\infty\binom{\frac1n-1}{2m}\binom{\frac12}{m+1}t_2^{2m}\,,
\end{equation}
where $t_2=\tan2\theta$.
One can show that this sum is the hypergeometric equation
\begin{equation}
f=\vphantom{a}_2F_1\left(1-\frac1{2n},\frac{n-1}{2n},2,-t_2^2\right)\,,
\end{equation}
or, for $n=-3$,
\begin{equation}
f=\vphantom{a}_2F_1\left(\frac76,\frac23,2,-t_2^2\right)\,,
\end{equation}
which can be easily evaluated numerically.
Note that for small angles this yields $f=1$ as expected.

\begin{figure*}
\centering
\includegraphics[width=0.49\textwidth]{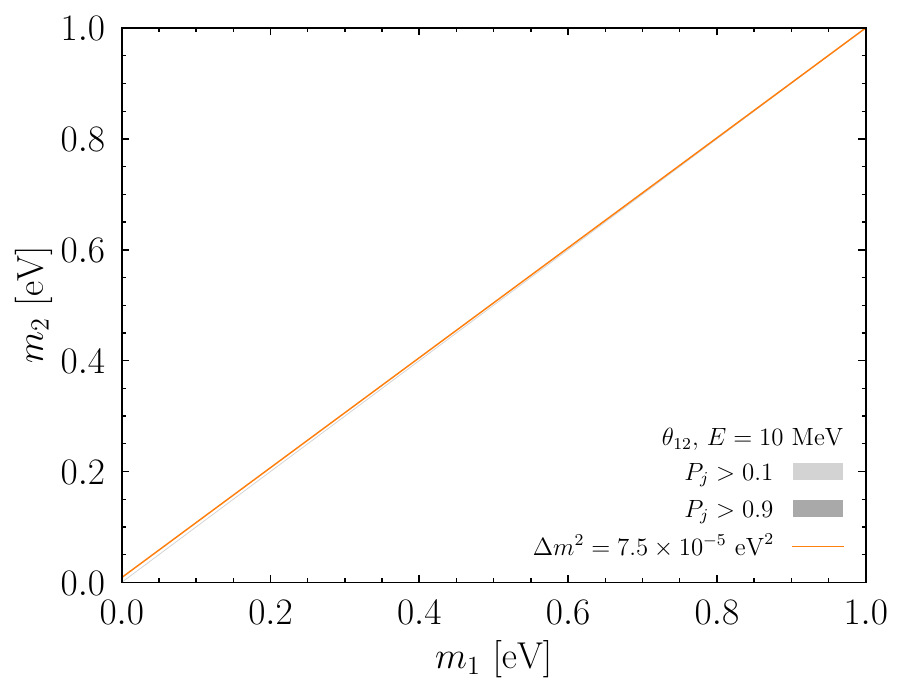}
\includegraphics[width=0.49\textwidth]{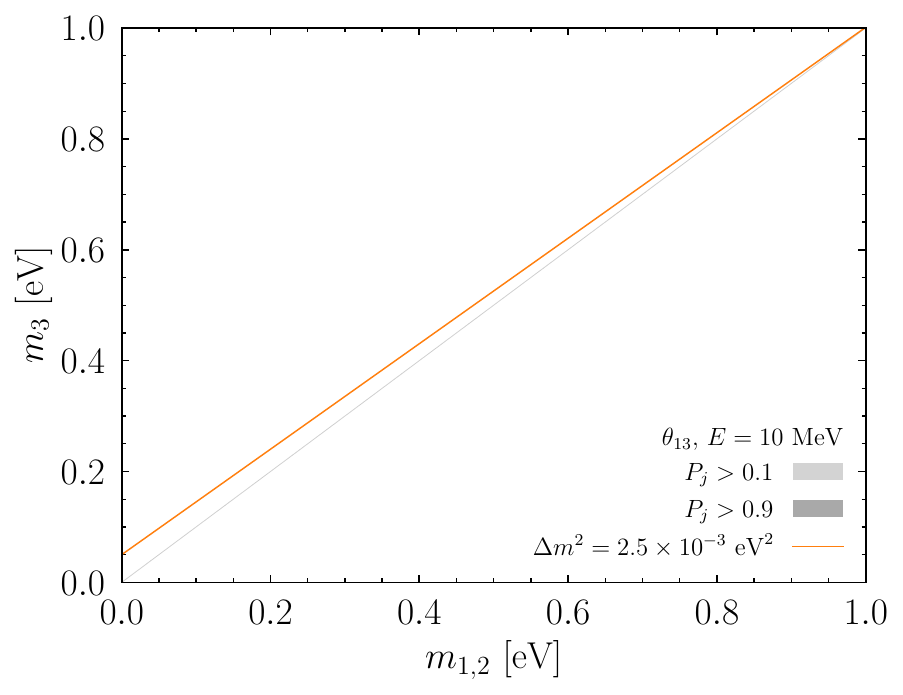}
\includegraphics[width=0.49\textwidth]{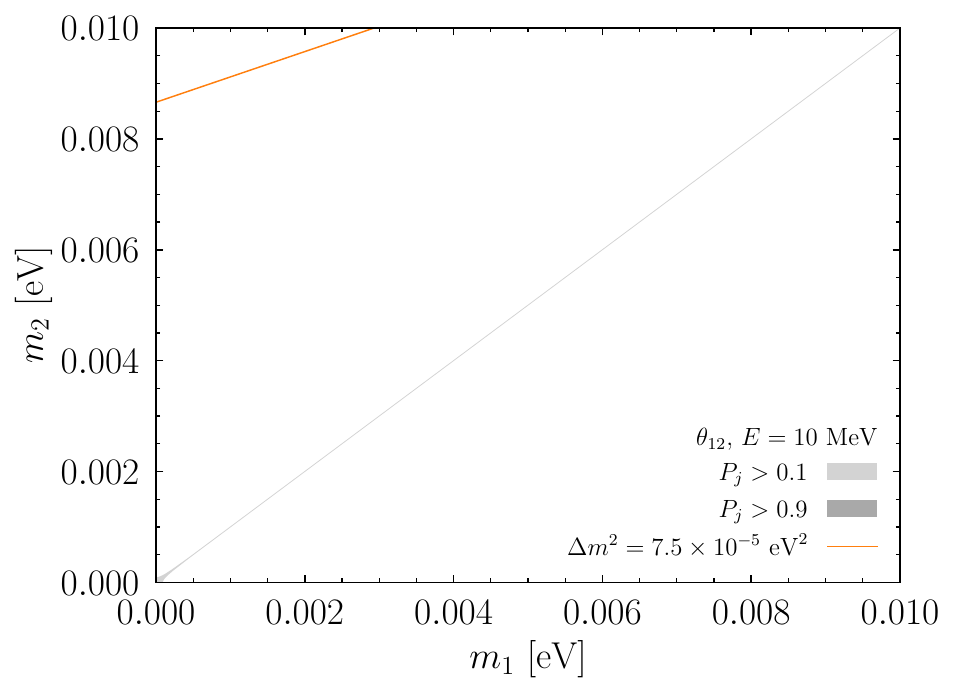}
\includegraphics[width=0.49\textwidth]{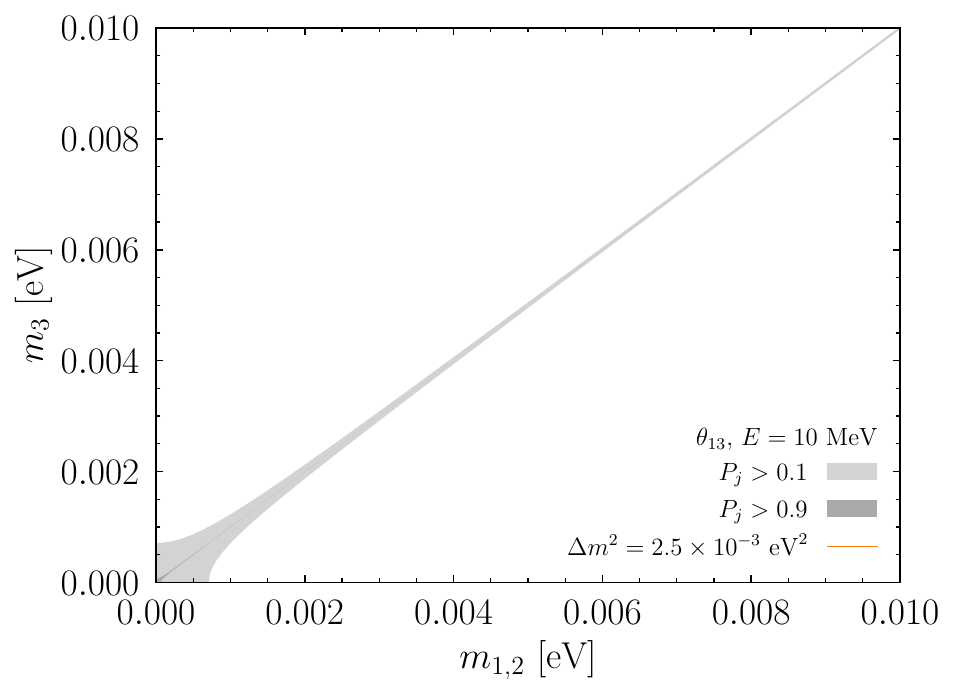}
\caption{The regions in $m_i$ parameter space at $E=10$ MeV where the $\Delta m^2$'s are small enough such that there are significant jump probabilities for the relevant mixing angle: \textbf{left}: solar, \textbf{right}: atmospheric splitting with the reactor angle.
The \textbf{lower} panels are zoomed in on the smallest masses.
No appreciable regions of parameter space here have $P_j>0.9$.
The orange lines are the actual $\Delta m^2$'s for the associated angles.}
\label{fig:10}
\end{figure*}

\begin{figure*}
\centering
\includegraphics[width=0.49\textwidth]{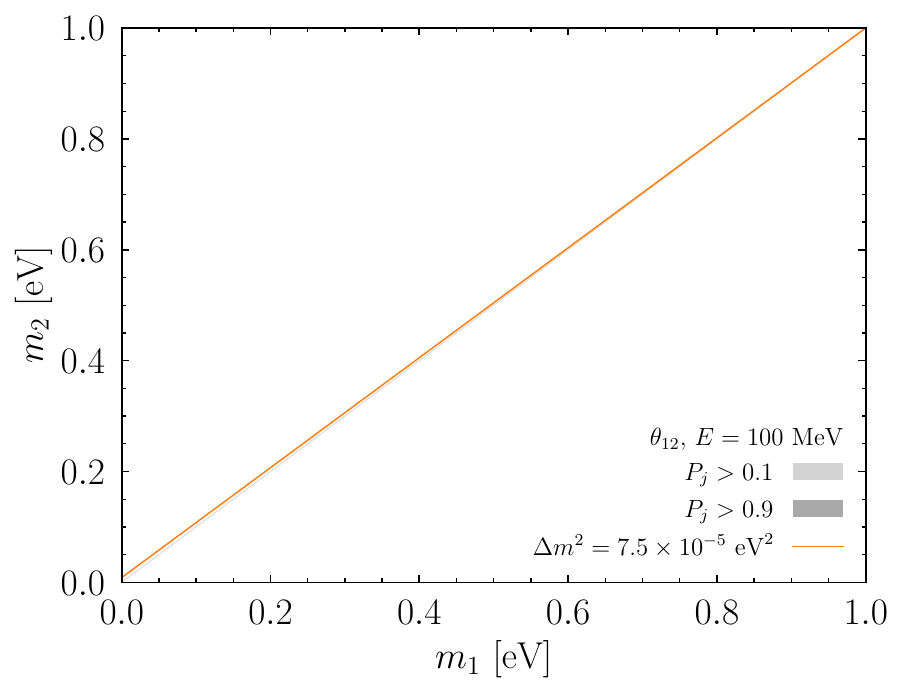}
\includegraphics[width=0.49\textwidth]{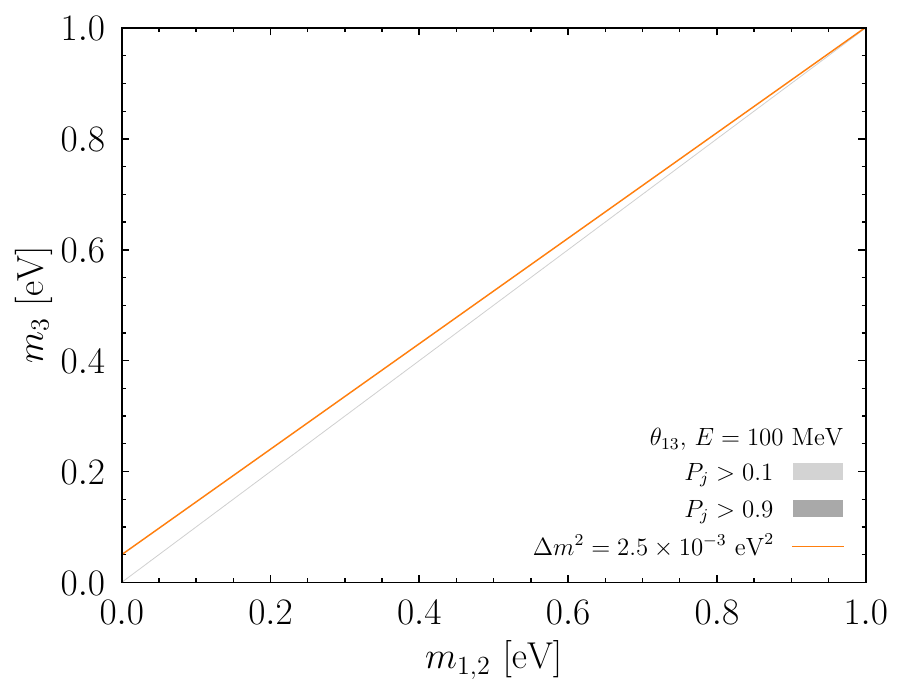}
\includegraphics[width=0.49\textwidth]{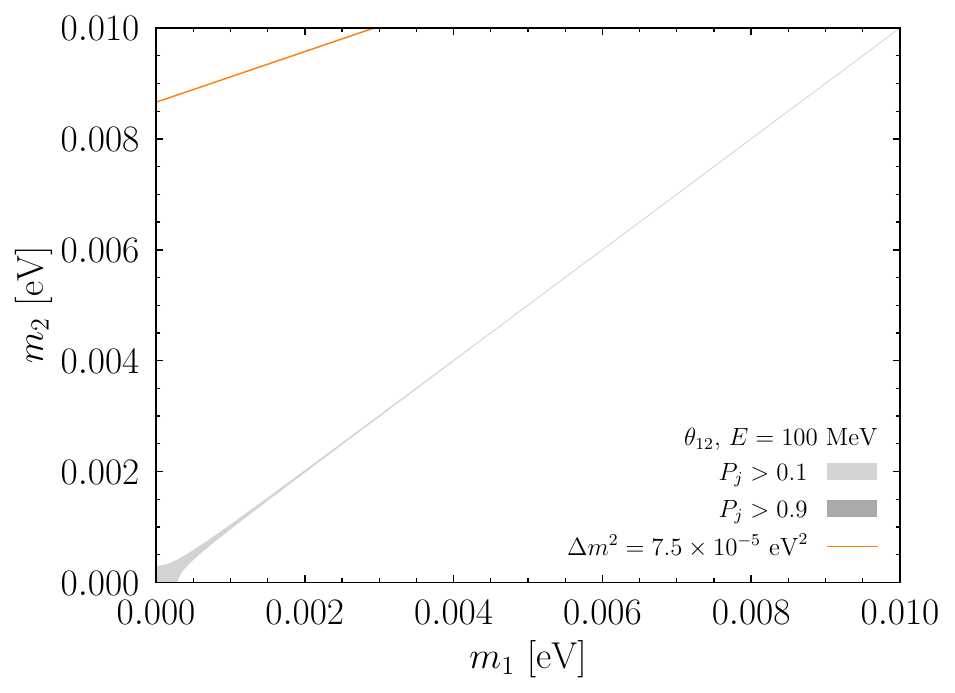}
\includegraphics[width=0.49\textwidth]{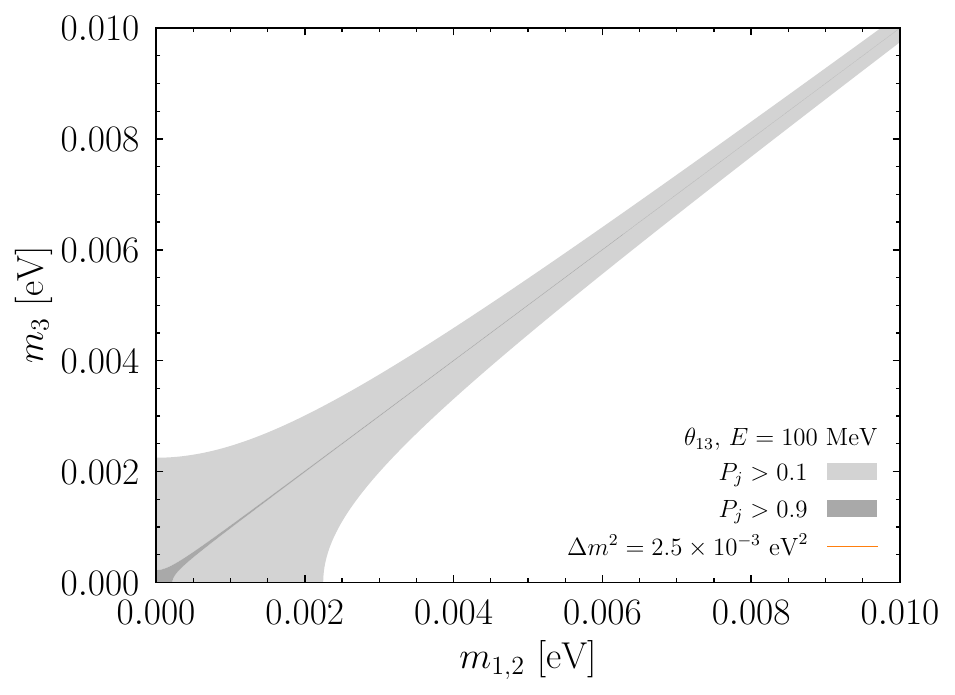}
\caption{The same as fig.~\ref{fig:10} except at $E=100$ MeV.}
\label{fig:100}
\end{figure*}

\begin{figure}
\centering
\includegraphics[width=0.49\textwidth]{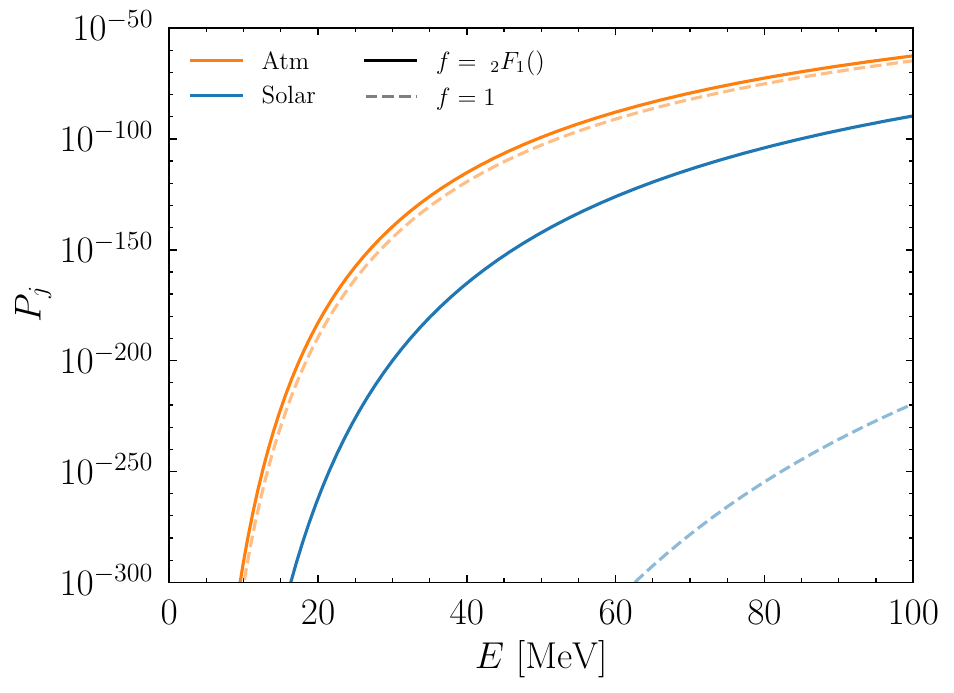}
\caption{The jump probability for the two standard mass splittings as a function of energy.
The solid curves use the correct form for $f$ while the dashed curves approximate $f=1$.}
\label{fig:standard}
\end{figure}

\subsubsection{Three-Flavor Picture}
We are in a three-flavor picture, however, which complicates things somewhat.
Since the formulas are derived in a two-flavor picture and involve an integrand of the form
\begin{equation}
\sqrt{(a-\Delta m^2\cos2\theta)^2+(\Delta m^2\sin2\theta)^2}\,,
\end{equation}
which is the the typical matter correction factor, we know how to best adjust these for a three-flavor picture, see \cite{Denton:2019yiw}, where $a=2EV_{\rm CC}$ quantifies the matter effect.

For the 21 sector, the factor is
\begin{equation}
\sqrt{(c_{13}^2a-\Delta m^2_{21}c_{212})^2+(\Delta m^2_{21}s_{212})^2}\,,
\end{equation}
and the other factor is most likely
\begin{equation}
\sqrt{(a-\Delta m^2_{e}c_{213})^2+(\Delta m^2_{ee}s_{213})^2}\,,
\end{equation}
where $s_{2ij}=\sin(2\theta_{ij})$, $c_{2ij}=\cos(2\theta_{ij})$, and
\begin{equation}
\Delta m^2_{ee}=c_{12}^2\Delta m^2_{31}+s_{12}^2\Delta m^2_{32}\,,
\end{equation}
see also \cite{Nunokawa:2005nx,Parke:2016joa}.

So the correct angle for the jump probability between 3 and 1 or 3 and 2 (depending on the hierarchy and neutrinos vs.~antineutrinos) is $\theta_{13}$ and the correct $\Delta m^2$ is probably $\Delta m^2_{ee}$.
For the other jump, between 1 and 2, the angle is $\theta_{12}$ and the matter effect (wherever $G_F$ shows up) should be multiplied by $c_{13}^2$.

For arbitrary $m_i$, these two-flavor approximations may not hold, and thus the corresponding angles may not be correct, but they should set the approximate scale which we will show is safe by many orders of magnitude.

\subsubsection{Numerical Studies}
We now numerically verify that these jump probabilities are small, using the known mixing angles.
First, we note that for $\theta=8.5^\circ$ and $34^\circ$ we get $f=0.97$ and $0.41$, respectively, thus the $f$ correction cannot be ignored.
Second, we let the masses vary across a range of $[0,1]$ eV and highlight the regions with noticeable jump probabilities as well as the expected regions if the known relevant $\Delta m^2$'s are to be satisfied.

We show the regions of $P_j\ge0.1$ and $P_j\ge0.9$ for different values of the two relevant masses for each relevant angle: $\theta_{12}$ and $\theta_{13}$, along with the zoomed-in region, at $E=10$ MeV, in fig.~\ref{fig:10}.
The same study is repeated at $E=100$ MeV in fig.~\ref{fig:100}, although this is just included to show the energy scaling; for the time delay calculation, the contribution to the information comes dominantly for lower-energy neutrinos.

We see that, except for a tiny sliver of parameter space, the effect is completely negligible.
It only begins to become relevant for energies higher than are relevant for the time delay effect and at masses that are very small and much closer together than anticipated from the $\Delta m^2$'s.
A new physics scenario involving masses this close together in the galactic center that then returns to the measured oscillation parameters in our solar system would be quite finely tuned.
In addition, the only region where a physical $\Delta m^2$, indicated by the orange lines, begins to become close to a noticeable jump probability is for masses much larger than consistent with cosmology or KATRIN.

Finally, one can also check that the jump probability is negligible for either mass splitting and for any relevant energy at the known $\Delta m^2$'s and mixing angles, as shown in fig.~\ref{fig:standard}.
We also see that while $f=1$ is a good approximation for a $\theta_{13}$ driven jump probability, it is not for a $\theta_{12}$ driven jump probability.

\subsection{Supernova Simulation}
A wide variety of SN simulations exist predicting a range of SN neutrino fluxes, see e.g.~\cite{Rampp:2002bq,Woosley:2007as,Nakazato:2012qf,OConnor:2014sgn,Sukhbold:2015wba,Mirizzi:2015eza,Vartanyan:2018iah,Skinner:2018iti,Warren:2019lgb,Kuroda:2020pta,Burrows:2020qrp,Zha:2021fbi}.
While there is no guarantee that any one model is correct or that reality even exists within the window of models, next generation experiments will have vastly improved capabilities for constraining the SN neutrino flux, see e.g.~\cite{Hyper-Kamiokande:2021frf}, provided that a SN is detected.

For our numerical results, we leverage \verb1SNEWPY1\footnote{A customized version of \texttt{SNEWPY} supporting arbitrary mass scenarios and time delay effects, is available at: \url{https://github.com/yveskini/snewpy}.} \cite{SNEWS:2021ewj} and \verb1SNOwGLoBES1 \cite{2021ascl.soft09019S} for the detection.
We take as our fiducial SN model the 27 $M_\odot$ explosion from \cite{Mirizzi:2015eza}.
This sets the size of the neutronization burst and the overall statistics.
We consider a typical distance to the SN of 10~kpc.

\section{Time Delay Features}
\label{sec:time delay}
Now that we have a clear picture of the physics inside the SN, we address the major effect in play to probe the masses of neutrinos: time delay due to massive neutrinos.

The time delay relative to traveling at the speed of light for neutrino mass state $i$ is
\begin{equation}
\Delta t_i(E)=t_{\nu_i}-t_c=D\left(\frac 1{v_i}-1\right)\simeq\frac D2\left(\frac{m_i}E\right)^2\,,
\end{equation}
where $D$ is the distance to the SN and $E$ is the neutrino energy.
This is then implemented in eqs.~\ref{eq:Phinue arbitrary}-\ref{eq:Phinuxbar arbitrary} by making the following changes throughout those equations:
\begin{equation}
|U_{\alpha i}|^2\Phi_{\nu_\beta}^0(E,t)\to|U_{\alpha i}|^2\Phi_{\nu_\beta}^0(E,t-\Delta t_i(E))\,,
\end{equation}
where we note that the mass state that should be applied for a given time delay depends on the prefactor of the PMNS matrix that is applied.
This makes it manifest why it is essential to not apply unitarity conditions to rewrite the flavor transformations into simpler looking expressions.

Having a time delay effect does not necessarily lead to an easily identifiable signal without a sharp feature in the spectrum.
That is, ideally the spectrum will suddenly and dramatically change due to some large-scale feature of the explosion.
We explore several such possibilities throughout this section and perform numerical studies for three of them in section \ref{sec:sensitivities}.
We note that while the exact details of these features may have sizable uncertainties in the magnitude and shape of the effect, they are unlikely to be degenerate with the time delay signature: a shifting of the low energy events into later times.

\subsection{Neutronization Burst}
It became clear fairly early in the history of SN simulations that an early peak of $\nu_e$ neutrinos is expected when the shock wave is first allowed to pass through the neutrinosphere \cite{Rampp:2000ws,Liebendoerfer:2000cq,Thompson:2002mw,Liebendoerfer:2002ny}, see also \cite{Kachelriess:2004ds}.
The burst lasts $\sim25$ ms and, while it does not contain the majority of the detectable SN neutrinos, its sharp features both turning on and turning off, as well as its dependence on the oscillation picture, make it an important diagnostic tool for extracting information about neutrinos from SN.
The neutronization burst is likely a signature of all SN, although its exact features do vary from one simulation to the next.
For our numerical studies, when focusing on the neutronization burst, we consider a timing window of 1 s starting from the bounce time in the simulation.

\subsection{QCD Phase Transition}
A QCD phase transition, i.e.,~a quark hadron phase transition, may occur in SN.
If so, the SN explosion includes a first-order phase transition from nuclear matter to quark matter.
This can produce a new source of energy to drive a second shock wave restarting a stalled shock.
See \cite{Sagert:2008ka,Fischer:2010wp,Fischer:2017lag,Zha:2021fbi,Fischer:2021tvv,Kuroda:2021eiv,Bauswein:2022vtq,Jakobus:2022ucs,Lin:2022lck,Pitik:2022jjh} for various discussions of this phenomenon.

In practice, this leads to a noticeable sharp feature in the spectrum that can likely be detected at high significance, depending on its height.
The sensitivity to neutrino masses from this sharp feature is independent of when postbounce it happens.
To ensure that we include the full delay effect, we include a window of width 40 ms focused on the peak and immediately after it.

\begin{figure*}
\centering
\includegraphics[width=0.49\textwidth]{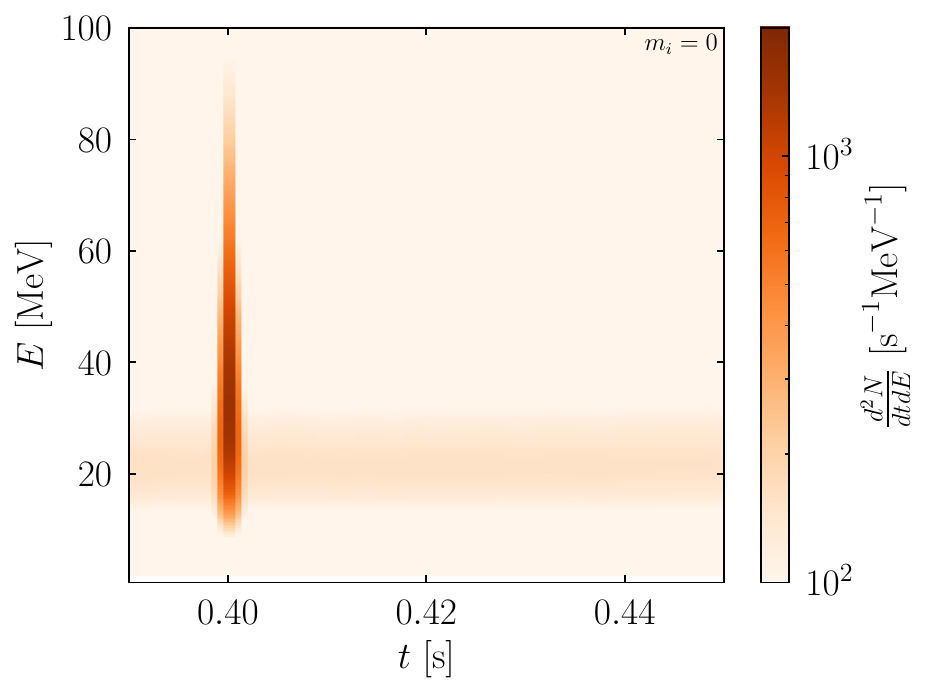}
\includegraphics[width=0.49\textwidth]{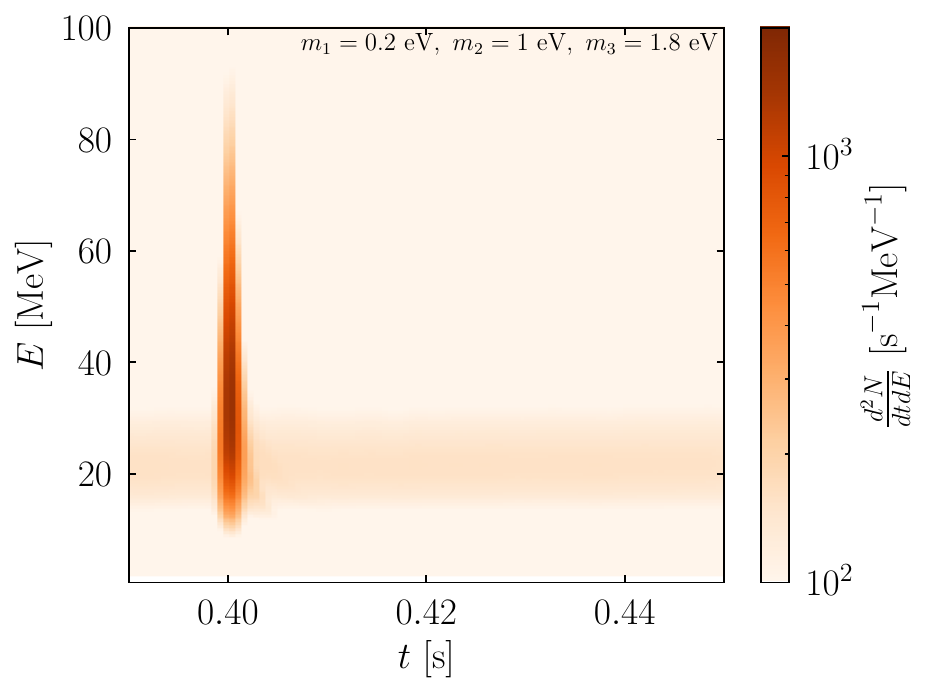}
\includegraphics[width=0.49\textwidth]{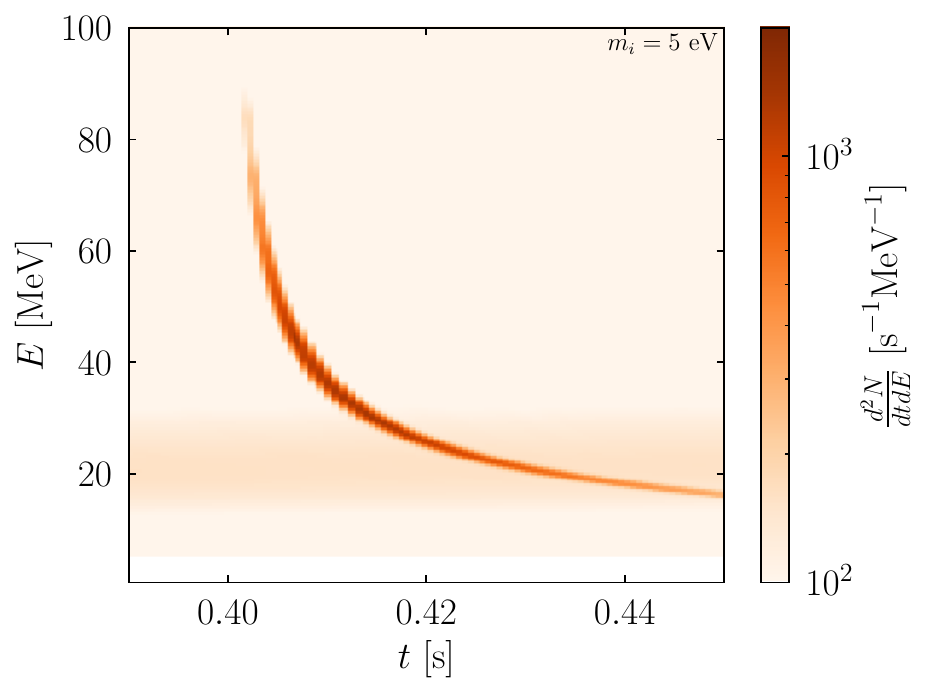}
\caption{The expected event rates for a SN at 10 kpc detected by JUNO as a function of energy and time for three different neutrino mass configurations around the QCD phase transition.
In the \textbf{top left} panel the neutrinos are massless, in the \textbf{top right} panel the neutrinos are in a varied high-mass scenario (see subsection \ref{sec:benchmarks} and table \ref{tab:benchmarks}), and in the \textbf{bottom} panel the neutrinos are all quite heavy.
Note that in the top right panel, some of the low-energy events are somewhat delayed due to $m_3$ large, but the effect of the remaining events is smaller. Additional plots for neutronization burst and black hole formation are provided in Appendix \ref{sec:event rates appendix}.}
\label{fig:events QCD}
\end{figure*}

We visually show the impact of the time delay due to a QCD phase transition in fig.~\ref{fig:events QCD} for different neutrino masses.
Note that in the middle case we see the time delay is stronger for some events than for others due to the different masses; this feature potentially allows for the discrimination of the different individual mass states.
The effect is similar for the other features in a SN; this case is shown for clarity.

\subsection{Black Hole Formation}
While many SN end in neutron stars as the final state of the compact object, a significant fraction of SN result in a BH \cite{Fryer:1999mi,Sekiguchi:2004ba,Sekiguchi:2004tma,Zhang:2007nw,Fischer:2008rh,Moller:2018kpn,Ziegler:2022ivq,Suliga:2022ica}.
The process of forming a BH will lead to a fairly sharp truncation in the neutrino signal which can be leveraged for timing purposes to determine the neutrino masses, see also \cite{Beacom:2000qy}.

We parameterize BH formation by rescaling the flux, mean neutrino energy, and pinching parameter via a sigmoid function. This ensures that the evolution of these quantities during BH formation closely matches the numerical results presented in \cite{Gullin:2021hfv} and includes the realistic turn-off time for the neutrino flux which has a characteristic time $\mathcal O(2)$ ms.
As with the QCD phase transition, we include a window of width 40 ms around the BH formation time.

\subsection{SASI}
Another possible feature that gives rise to a time-dependent signature is the Standing Accretion Shock Instability (SASI) \cite{Blondin:2002sm,Foglizzo:2015dma,Lin:2019wwm}.
This leads to an oscillatory signature in the neutrino flux.
When combined with time delay, which depends on energy, it could be possible to extract information on neutrino masses.
We do not consider this effect in this study because the magnitude of this effect varies considerably from simulation to simulation, the effect tends to get largely washed out if the progenitor carries moderate or large angular momentum, and marginalizing over the uncertainties in the oscillatory signature will likely wash out the effect considerably.
Nonetheless, if a galactic SN neutrino signal is detected and if it carries evidence of SASI, a time delay analysis on that region of the data should also be performed.

\section{Detection}
\label{sec:detection}
A large number of current neutrino experiments have sensitivities to measuring a galactic SN significantly improved over what was present in 1987.
These include Super-Kamiokande \cite{Super-Kamiokande:2002weg,Super-Kamiokande:2022dsn,KamLAND:2024uia,Super-Kamiokande:2024bsv,Super-Kamiokande:2024pmv}, IceCube \cite{IceCube:2011cwc,Kopke:2017req}, and NOvA \cite{NOvA:2020dll}.
In the coming years, JUNO \cite{JUNO:2023dnp}, Hyper-Kamiokande \cite{Hyper-Kamiokande:2018ofw,Migenda:2019xbm,Hyper-Kamiokande:2021frf}, and DUNE \cite{DUNE:2020ypp,DUNE:2020zfm} will also come online, each with its own advantages, leading to the expectation of high statistics and broad coverage in terms of physics to be extracted from a galactic SN, see also \cite{Scholberg:2017czd}.

\subsection{JUNO}
In this article, we will focus on detection with JUNO \cite{JUNO:2021vlw,JUNO:2023dnp}.
Even though the detector volume of JUNO will be smaller than some other large-volume neutrino detectors, it is expected to come online before other next-generation experiments DUNE and HK.
In addition, due to the usage of liquid scintillator and state-of-the-art photomultiplier tubes, large overburden, and good radiopurity, JUNO is expected to achieve fairly low thresholds.
Since the time delay effect scales like $E^{-2}$, probing lower energy neutrinos results in a greater advantage for determining neutrino masses.
In addition, this will reduce the impact due to the uncertainty in the nature of the SN neutrino flux itself by measuring the time delay effect across a range of energies where the dominant contribution to the event rate comes in the energy range $5\lesssim E/$MeV$\lesssim50$.
We implement JUNO's detection capabilities as described in \verb1SNowGLoBES1 \cite{2021ascl.soft09019S} including cross sections and experimental details which also accounts for neutrino energy smearing for all detection channels.
We also find that JUNO will likely have excellent timing resolution $\lesssim1$ ns, better than needed for typical SN, given the $<10$ cm spatial resolution \cite{Qian:2021vnh}.

\subsection{Cross Sections}
Beyond the advantage of lower thresholds, JUNO will also be sensitive to a number of different cross section channels including inverse beta decay (IBD), electron elastic scattering, $^{12}$C, $^{13}$C, and neutral current, with decent discrimination among them \cite{2021ascl.soft09019S,Lu:2014zma,JUNO:2023dnp}.
We optimistically assume that JUNO can differentiate all the relevant channels, although we find that this approximation does not affect our results too much as most of the statistics come from IBD.

Proton elastic scattering (pES) also contributes a significant amount $\mathcal O(10\%)$ to the event rate, and the cross section is thresholdless \cite{Lu:2016ipr}.
The cross section is, however, rather uncertain at sub-GeV energies \cite{Chauhan:2022wgj} and while the JUNO thresholds for proton detection may be at the 0.2 MeV level, in practice this corresponds to $\gtrsim10$ MeV neutrino energies, at which point IBD still performs much better for time delay.
Furthermore, pES experiences quenching which further reduces the low energy signature and is an additional source of uncertainty.
For these reasons, we do not include pES in our analysis, but doing so would enhance the sensitivity to neutrino masses somewhat.

\section{Sensitivities}
\label{sec:sensitivities}
In this section we numerically compute the sensitivity to disfavor certain combinations of neutrino masses given a detection of a galactic SN at 10 kpc.
We first discuss some benchmarks for the true neutrino masses to be considered, and then we numerically compute the preferred regions assuming a Poisson log-likelihood ratio.

\subsection{Benchmarks}
\label{sec:benchmarks}
In order to compute sensitivities, we consider a variety of benchmarks derived from existing determinations of neutrinos masses.

The cleanest and most robust information we have about neutrino masses comes from oscillations, specifically the fact that $\Delta m^2_{21}=+7.4\e{-5}$ eV$^2$ and $|\Delta m^2_{31}|=2.5\e{-3}$ eV$^2$ with uncertainties at the $\sim3\%$ and $\sim1\%$, respectively.
While the sign of $\Delta m^2_{31}$ remains undetermined\footnote{Several years ago, it did appear that the data was hinting that the sign was positive, but the data is now less clear, see e.g.~\cite{Esteban:2024eli}.}, it is strongly expected that this will be quite clear in the future oscillation picture with data from T2K, NOvA, Daya Bay, RENO, JUNO, IceCube, SK, KM3NeT, DUNE, and HK contributing \cite{T2K:2023smv,NOvA:2004blv,DayaBay:2022orm,RENO:2018dro,NOvA:2021nfi,DUNE:2020ypp,DUNE:2022aul,Denton:2022een,Nunokawa:2005nx,Parke:2024xre}.
This is known as the atmospheric mass ordering and is called the normal ordering (NO) for $\Delta m^2_{31}>0$ and the inverted ordering (IO) for $\Delta m^2_{31}<0$.
With this existing and future oscillation information, only one more measurement is needed to determine the absolute mass scale, which could come from a variety of directions.

Cosmological measurements constrain the quantity $\sum_im_i$.
The data indicates that this quantity is consistent with zero (actually slightly preferring negative values) \cite{Planck:2018vyg,DESI:2024mwx,Craig:2024tky,Jiang:2024viw} and is putting pressure on the inverted mass ordering (the case where the neutrino that is least $\nu_e$ is the lightest), which has a larger minimum sum of the neutrino masses.
In fact, the data is also somewhat disfavoring the normal ordering as well, at low significance.
While future data is expected to have good sensitivity to measuring this quantity, the current situation with the data, as well as tensions in cosmology \cite{Abdalla:2022yfr,Poudou:2025qcx}, is somewhat complicated.
In light of this, we use the constraint from Planck alone (TT, TE, EE, lowE, and lensing) as one of our benchmarks $\sum_im_i<0.24$ eV \cite{Planck:2018vyg}.

A cleaner but less competitive measurement comes from KATRIN looking at the endpoint of the tritium spectrum \cite{Katrin:2024tvg}\footnote{See also Project 8 \cite{Project8:2022wqh} as well as the electron capture proposals HOLMES \cite{HOLMES:2019ykt} and ECHo \cite{Gastaldo:2013wha}.} and is expected to gain another factor of $\sim2$ in precision, although this will still not be enough to be competitive with the cosmological constraint.
Other probes are possible as well such as neutrinoless double beta decay provided that neutrinos have Majorana mass terms and the light neutrino exchange model is valid.
It may also be possible to get directly at the three separate neutrino masses in other ways such as via a precise measurement of the cosmic neutrino background \cite{Long:2014zva} via an experiment such as PTOLEMY \cite{PTOLEMY:2018jst}, although this faces significant experimental challenges for even a detection, let along a separate mass determination \cite{Cheipesh:2021fmg,Nussinov:2021zrj,PTOLEMY:2022ldz}.

Combining this, our benchmark points\footnote{It is important to note that, while our benchmarks assume normal and inverted mass orderings for simplicity, we compute spectra and event rates for all possible orderings and perform a statistical comparison among them.
Our results will be similar for different true mass orderings.}, shown in table \ref{tab:benchmarks}, are either the minimum masses allowed by oscillations, or the maximum masses allowed by other data sets.
We also include one high mass scenario that does not take into account the measured $\Delta m^2_{ij}$'s, that could be due to new physics.
We have also investigated other scenarios such as the upper limit on the masses from various cosmological data sets including CMB, BAO, and the latest BAO updates from DES and DESI.
We found that the constraints in those cases were quite similar to the lowest possible masses.

\begin{table}
\centering
\caption{The benchmark points for a variety of scenarios.
Some are the lightest possible masses given oscillation data alone in either mass ordering.
The heaviest masses allowed by KATRIN which is the same for either mass ordering, and the high mass scenario is one that requires new physics (see section \ref{sec:spatial neutrino masses}) which can lead to different masses elsewhere in the galaxy.}
\label{tab:benchmarks}
\begin{tabular}{l|c|c|c}
Scenario & $m_1$ [eV] & $m_2$ [eV] & $m_3$ [eV] \\\hline
Lightest by oscs: NO & 0.0 & 0.009 & 0.050\\
Lightest by oscs: IO & 0.050 & 0.051 & 0.0\\
Heaviest by Planck: NO & 0.075 & 0.075 & 0.090\\
Heaviest by Planck: IO & 0.085 & 0.086 & 0.069\\
Heaviest by KATRIN & 0.45 & 0.45 & 0.45\\
High mass scenario & 0.2 & 1.0 & 1.8
\end{tabular}
\end{table}

\begin{figure*}
\centering
\includegraphics[width=0.49\textwidth]{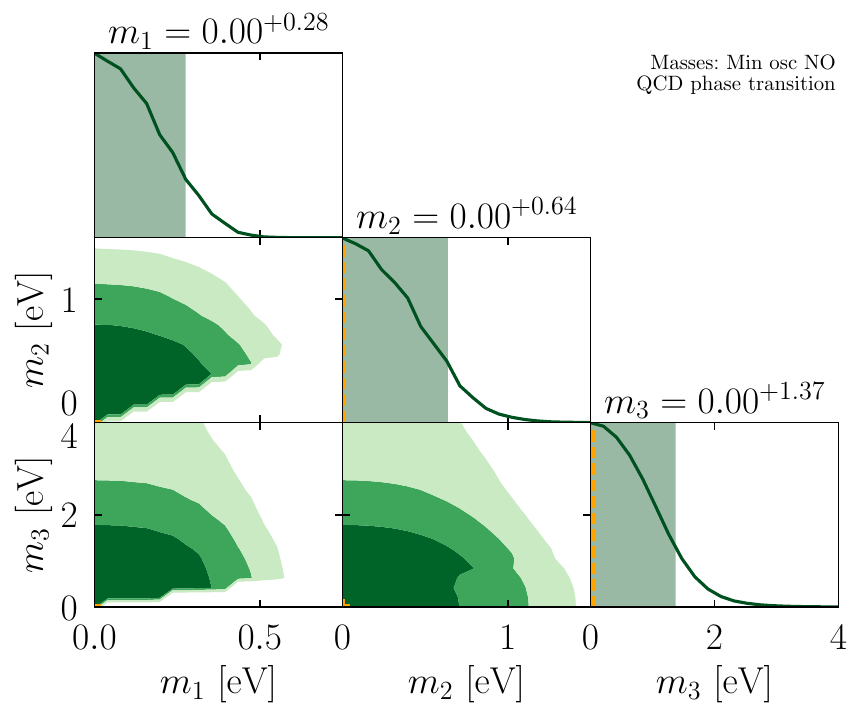}
\includegraphics[width=0.49\textwidth]{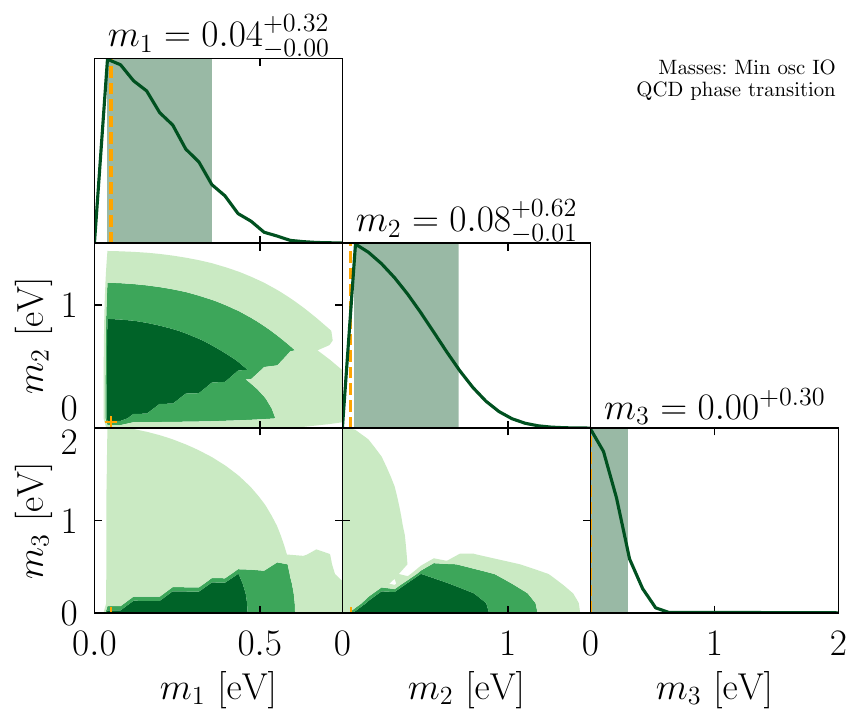}
\includegraphics[width=0.49\textwidth]{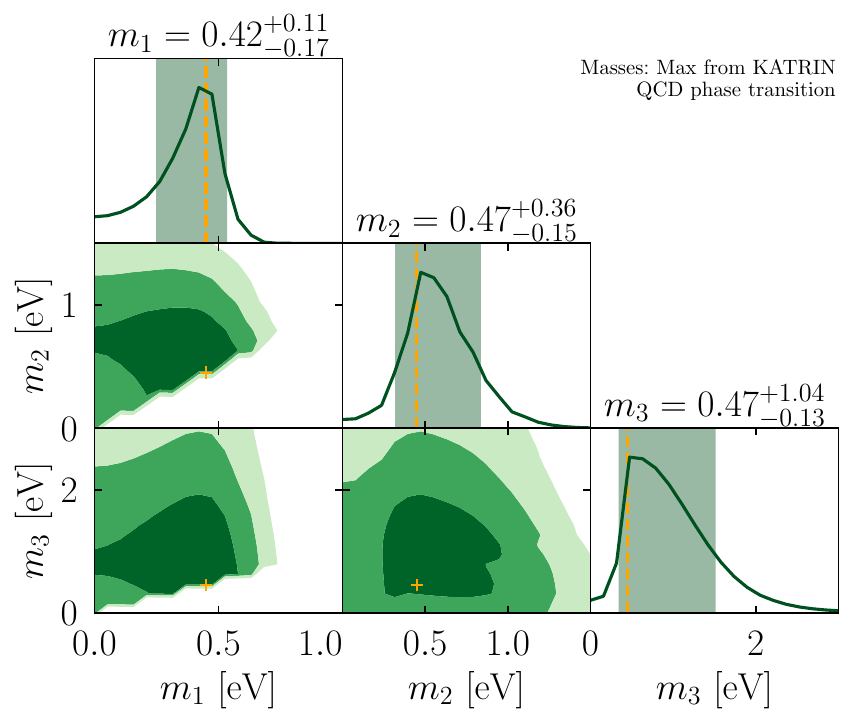}
\includegraphics[width=0.49\textwidth]{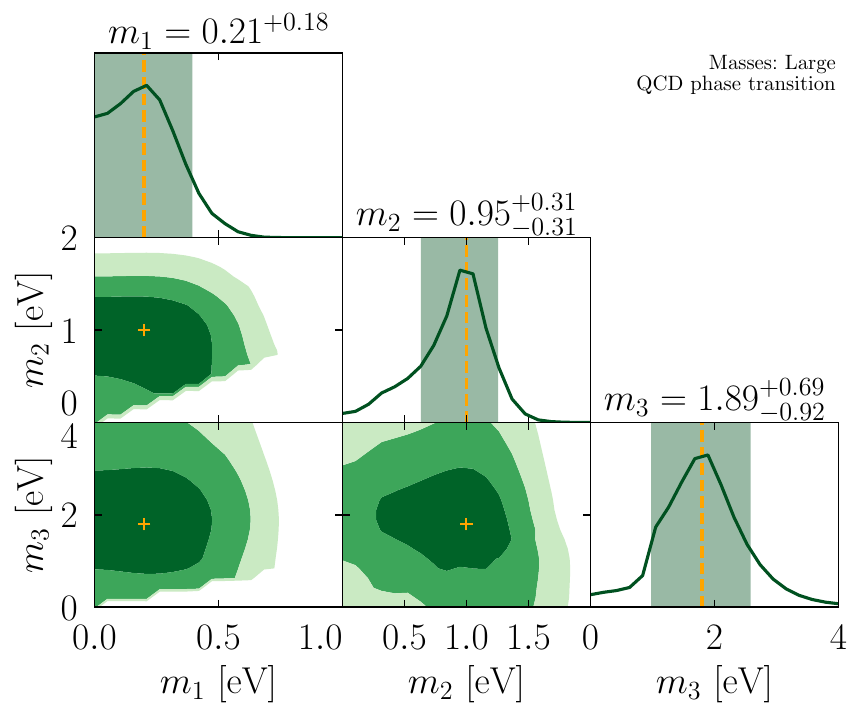}
\caption{The expected preferred regions given a detection of a SN at 10 kpc by JUNO under four different hypotheses about the true neutrino masses (see table \ref{tab:benchmarks}).
Note that in this case, \emph{only} information from the QCD phase transition is used to determine the neutrino masses.
The contours are drawn at 1, 2, and 3 $\sigma$ assuming 2 dofs and Wilks' theorem, and the shaded regions correspond to the 1 $\sigma$ region for 1 dof.
The pluses and vertical dashed lines represent the true values.}
\label{fig:corner QCD}
\end{figure*}

We again reiterate our above discussion around eq.~\ref{eq:normalcy} that we determine which mass state is which by the magnitude of the elements of the electron neutrino row of the PMNS matrix.
Specifically, we take $|U_{e1}|^2=0.68$, $|U_{e2}|^2=0.30$, and $|U_{e3}|^2=0.02$ regardless of the masses being considered.

\begin{figure*}
\centering
\includegraphics[width=0.49\textwidth]{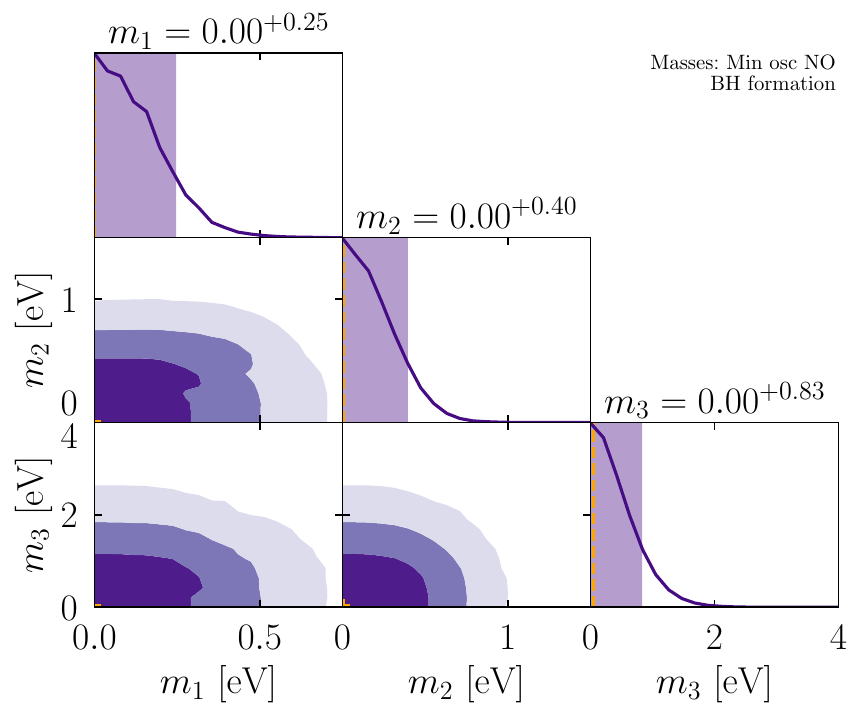}
\includegraphics[width=0.49\textwidth]{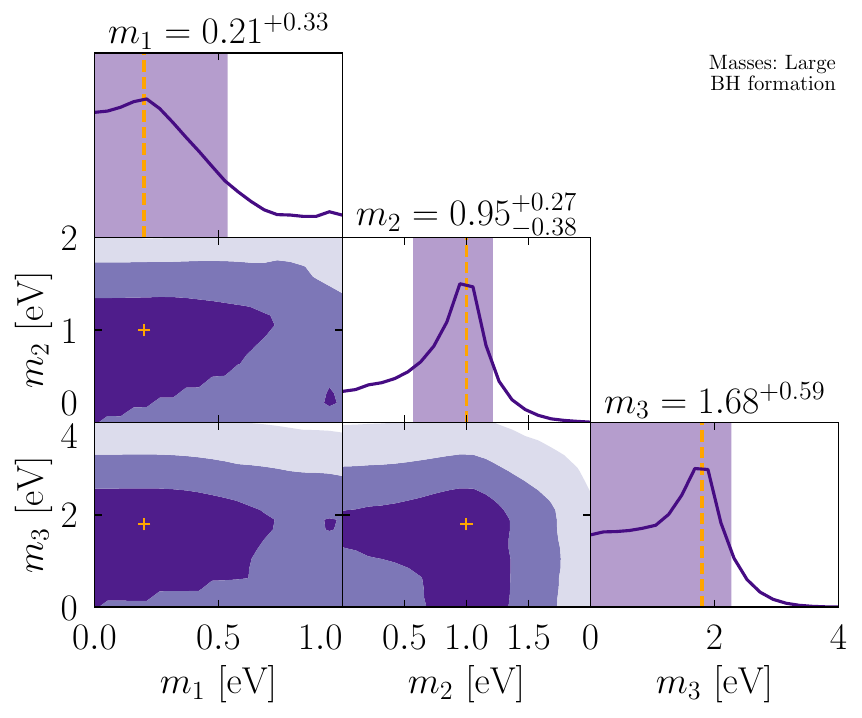}
\caption{The same as fig.~\ref{fig:corner QCD} but for the BH case.}
\label{fig:corner BH}
\end{figure*}

\subsection{Numerical Results}
Here we show our numerical results.
We compute a $\chi^2$ test statistic expressed as the log-likelihood ratio of Poisson probability density functions to account for low-statistic bins.
We assume Wilks' theorem \cite{Wilks:1938dza} to compute our confidence intervals with two degrees of freedom for the 2D projections and one degree of freedom for the 1D projections. 
This approach may be conservative in estimating confidence intervals, see \cite{Lu:2014zma}.
We also minimize over the start time for each of the three features as this is the largest uncertainty.
Additional uncertainties due to the details of the SN itself seem to have a smaller effect \cite{Lu:2014zma}.

\begin{figure*}
\centering
\includegraphics[width=0.49\textwidth]{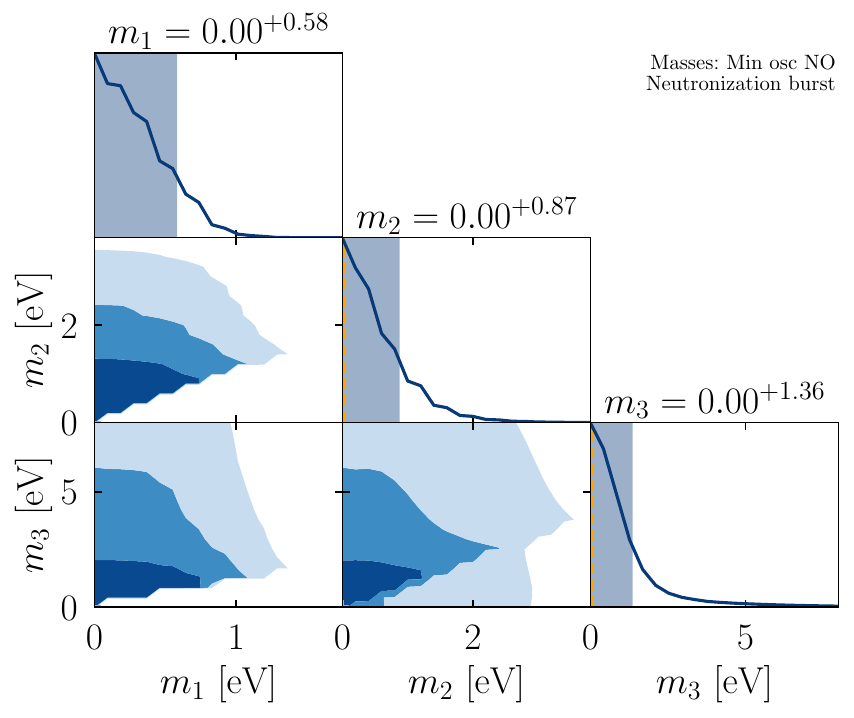}
\includegraphics[width=0.49\textwidth]{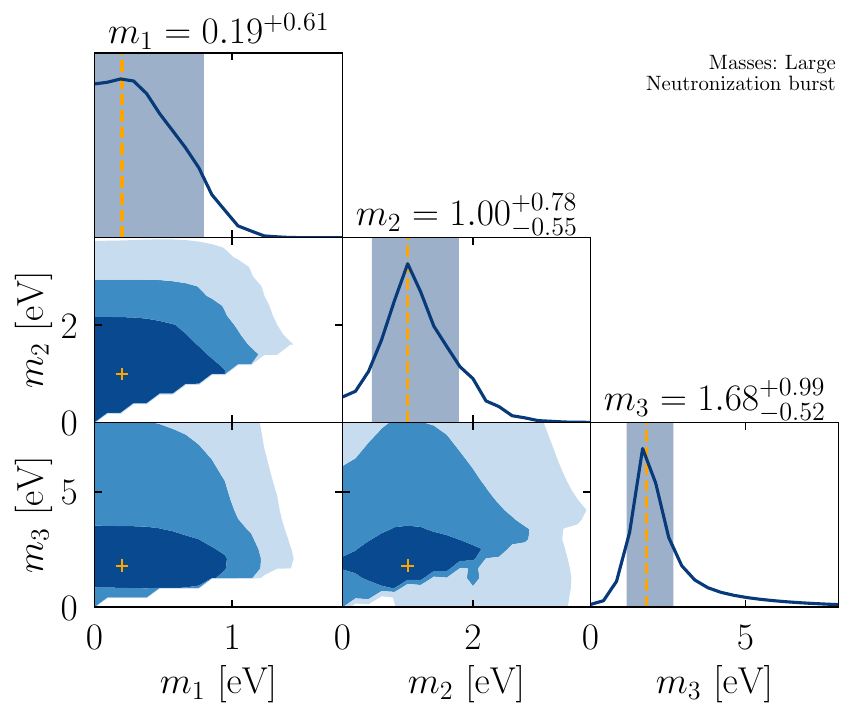}
\caption{The same as fig.~\ref{fig:corner QCD} but for the regular (neutronization burst) case.}
\label{fig:corner NB}
\end{figure*}

We show corner plots for the QCD phase transition case for various mass benchmarks in fig.~\ref{fig:corner QCD}.
We consider different physical effects that could be used to determine the neutrino masses: the neutronization burst, a QCD phase transition, and the formation of a BH.
We also consider different true values for the masses.
In figs.~\ref{fig:corner BH} and \ref{fig:corner NB} we also show some results for the BH case and the regular (neutronization burst) case.
Additional numerical results for all the benchmarks can be found in the appendix \ref{sec:sensitivities appendix}.
Our numerical results of all cases are also summarized in table \ref{tab:results}.

\begin{table*}
\centering
\caption{The $1\sigma$ sensitivities for each of the three mass states for each of the benchmarks and for each of the physical effects in the SN given a detection of a SN at 10 kpc by JUNO.
All masses are in eV.
See table \ref{tab:benchmarks} for the definitions of each benchmark.}
\label{tab:results}
\begin{tabular}{l|c|c|c||c|c|c||c|c|c}
& \multicolumn{3}{|c||}{\textbf{QCD}} & \multicolumn{3}{c||}{\textbf{BH}} & \multicolumn{3}{c}{\textbf{NB}}\\\hline
\textbf{Benchmarks} & $m_1$ & $m_2$ & $m_3$ & $m_1$ & $m_2$ & $m_3$ & $m_1$ & $m_2$ & $m_3$\\\hline
Oscs: NO & $<0.28$ & $<0.64$ & $<1.37$ & $<0.25$ & $<0.40$ & $<0.83$ & $<0.58$ & $<0.88$ & $<1.36$\\
Oscs: IO & [0.04, 0.36] & [0.07, 0.70] & $<0.30$ & [0.03, 0.32] & [0.01, 0.42] & $<0.30$ & [0.092, 0.612] & [0.13, 1.01] & $<0.52$ \\
Planck: NO & $<0.29$ & $<0.65$ & $<1.39$ & $<0.26$ & $<0.40$ & $<0.85$ & $<0.58$ & $<0.91$ & $<1.36$\\
Planck: IO & [0.04, 0.37] & [0.07, 0.72] & $<0.31$ & [0.03, 0.33] & [0.01, 0.42] & $<0.30$ & [0.09, 0.62] & [0.13, 1.01] & $<0.56$\\
KATRIN & [0.25, 0.54] & [0.32, 0.84] & [0.34, 1.52] & $<0.53$ & $<0.64$ & $<1.07$ & $<0.75$ & $<1.02$ & $<1.43$\\
High mass & $<0.40$ & [0.63, 1.26] & [0.98, 2.58] & $<0.54$ & [0.57, 1.21] & $<2.27$ & $<0.80$ & [0.45, 1.78] & [1.17, 2.68]
\end{tabular}
\end{table*}

\begin{figure}
\centering
\includegraphics[width=0.49\textwidth]{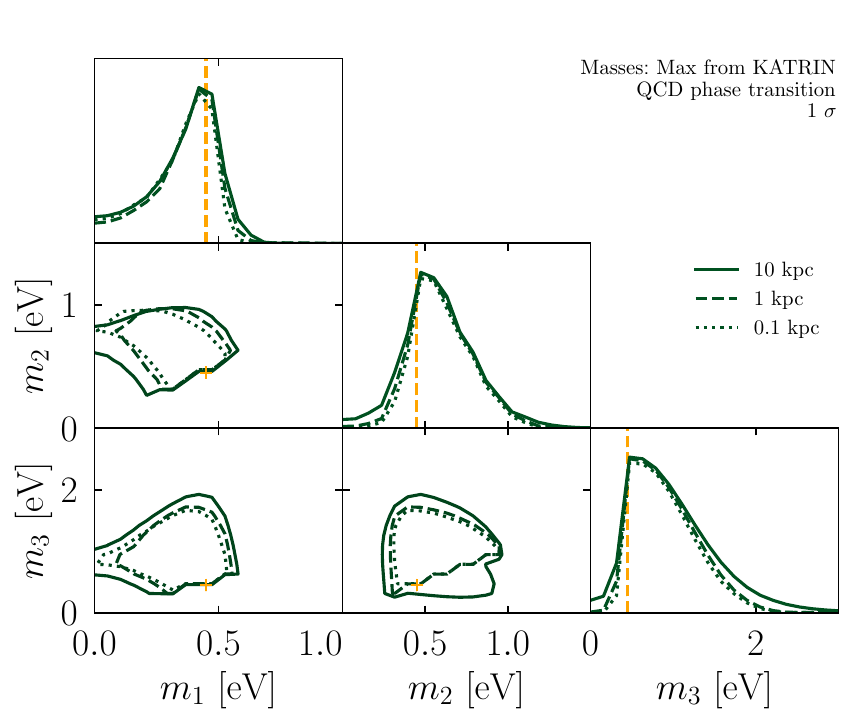}
\caption{The same as fig.~\ref{fig:corner QCD} for a QCD phase transition with masses at the upper limit from KATRIN for three different distances denoted by different line styles: the solid, dashed, and dotted curves represent distances of 10 kpc, 1 kpc, and 0.1 kpc, respectively.
The contours in the 2D panels are drawn for 1 $\sigma$.}
\label{fig:distance}
\end{figure}

Finally, we study the impact of distance on the precision of determining the neutrino mass states.
We consider the representative case of a QCD phase transition with neutrino masses at the limit from KATRIN, but other scenarios have similar results.
The results for three different distances, 0.1 kpc (roughly the distance to Betelgeuse), 1 kpc, and 10 kpc are shown in fig.~\ref{fig:distance}.
We see that closer SN generally result in tighter constraints, although for very light $m_1$ the sensitivity no longer improves at very short distances due to finite timing resolution.
We also see other smaller features such as that while the ordering $m_3>m_2$ cannot always be determined at 10 kpc, it can at 1 kpc or less, in this scenario.

\section{Discussion}
\label{sec:discussion}
The numerical studies have yielded many interesting and nontrivial results.
Perhaps the most compelling result found is the fact that in general there is information that can distinguish among the three different mass states in a galactic SN.
That is, depending on the true masses of neutrinos, each individual mass state can be uniquely measured.
The uniqueness statement is based on the fact that we assume that the PMNS matrix is determined.
Thus it is not surprising that the constraints on $m_3$ tend to be poorer because $|U_{e3}|$ is much smaller than $|U_{e1}|$ and $|U_{e2}|$.
In fact, we believe that any estimate of neutrino masses from a galactic SN should use the full description including the explicit contribution from each separate mass term as described in eqs.~\ref{eq:Phinue arbitrary}-\ref{eq:Phinuxbar arbitrary}.

In the numerical fits we can see some explicit information corresponding to the various mass orderings.
For example, in fig.~\ref{fig:corner QCD}, the 2D projections show that the scenario where $m_1>m_3$ in the true NO (or $m_3>m_1$ in the true IO) is disfavored at a significant level.
While some of this information comes from an incorrect time delay signature in propagation, the majority of it comes from the different MSW behavior past the resonances inside the SN.

We investigated the role of each individual cross section on the sensitivities.
While JUNO does benefit from some thresholdless processes allowing for lower energy measurements, and thus signatures of larger time delays, we found that the constraints are typically statistics dominated.

While we have focused on one detector, additional detections by DUNE, HK, and IceCube among others will provide additional information here which is expected to improve these constraints somewhat.

In this study we have assumed Wilks' theorem provides a robust statistical estimate of the confidence levels.
In reality several effects may alter this such as low statistics in the key regions of interest affected by time delay.
For a more careful treatment of these effects see e.g.~\cite{Pitik:2022jjh,Pompa:2022cxc}.

Finally, we comment that we only considered one physical process at a time: QCD phase transition, BH formation, and a neutronization burst.
While most SN are expected to have a neutronization burst, it is not guaranteed to exist, or rather to be large enough to be relevant.
That said, those SN with a sizable neutronization burst may well also experience a QCD phase transition and/or form a BH.
Thus additional information may exist to constrain neutrino masses from a single event.
As the underlying physics processes for each of these three events are unrelated, one can naively add the $\chi^2$ test statistics at each point in parameter space before the minimization over the other mass states to get the combined test statistic and the enhanced significance.

\section{Conclusions}
\label{sec:conclusions}
Determining the absolute mass scale of neutrinos is not possible via oscillation experiments and requires a novel approach.
Cosmological measurements of the CMB and large scale structure promise the most sensitivity, but data is currently weakly hinting at some incompatibility with oscillation data.
In light of this, we considered several alternative scenarios, either motivated by the constraints from the Planck satellite alone, the constraints from KATRIN here on Earth, or a large mass scenario as might be possible in various new physics scenarios discussed in section \ref{sec:spatial neutrino masses}.

While it has been known that a galactic SN can provide information about the masses of neutrinos that is complementary to that of oscillations, we have shown here that a galactic SN can actually identify each individual mass state.
This is partially due to the distinct signatures of mass orderings inside the SN affect the resonant behavior, but dominantly due to the fact that each mass state propagates at a different speed and these effects can be differentiated, see figs.~\ref{fig:events QCD}.

The time delay effect is cleanly identifiable from sharp features in the SN spectrum and we investigated three such features: the neutronization burst which is likely present in all SN, a QCD phase transition which may or may not be present in all SN, and the formation of a BH which happens in a subset of SN.

We computed numerical statistical studies to estimate JUNO's sensitivity to determine each individual mass state in each different physics scenarios for a variety of neutrino mass benchmarks.
We found that, consistent with the theoretical arguments presented, a detection of a galactic SN can measure the mass of each neutrino state, subject to statistical uncertainties.
If the masses are on the larger side, such as the limit from KATRIN, each mass can be measured.
And while the individual neutrino masses are similar in this case, if new physics indicates that neutrino masses are different elsewhere in the galaxy, as may be the case due to DM interactions, each separate mass state can be uniquely identified.

This opens the very real possibility for a galactic SN to go beyond anticipated measurements of neutrino mass-squared differences and the sum of the neutrino masses and provide an powerful orthogonal constraint on neutrino masses that could help us understand the nature of neutrino mass generation.

\begin{acknowledgments}
We thank Anna Suliga for many helpful discussions throughout this project.
PBD acknowledges support by the United States Department of Energy under Grant Contract No.~DE-SC0012704. YK acknowledges support from European
Research Council (ERC)  Consolidator Grant No.~865768 AEONS (PI Anna L.~Watts). The major part of the work was carried out on the HELIOS cluster exclusively on dedicated nodes funded via the abovementioned ERC CoG.
The figures were made with \texttt{python} \cite{10.5555/1593511} and \texttt{matplotlib} \cite{Hunter:2007}.
\end{acknowledgments}

\appendix

\section{Impact of Massive Neutrinos on Event Rates}
\label{sec:event rates appendix}
\begin{figure*}
\centering
\includegraphics[width=0.32\textwidth]{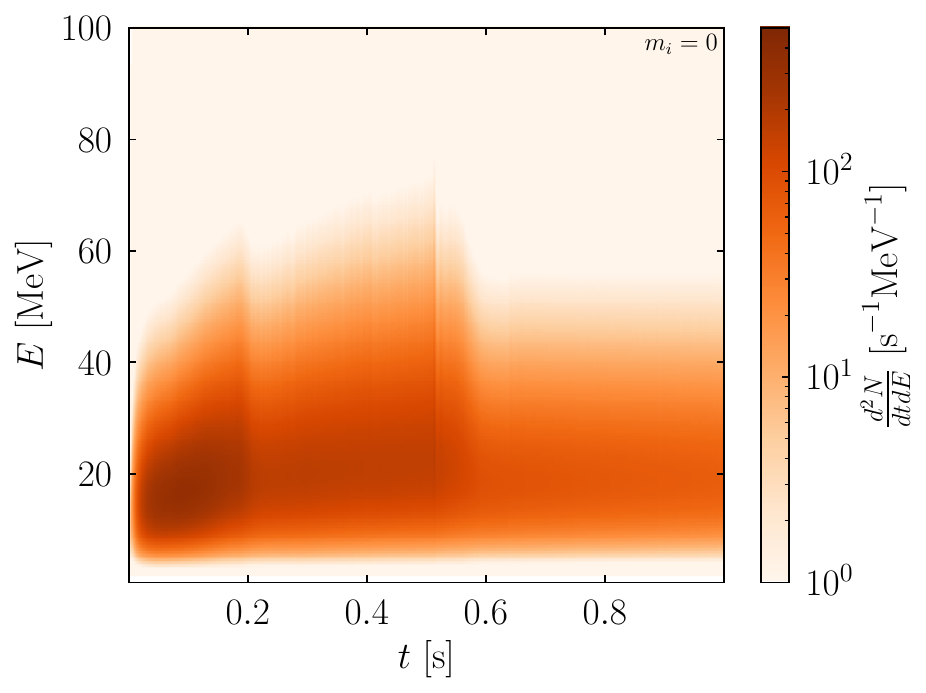}
\includegraphics[width=0.32\textwidth]{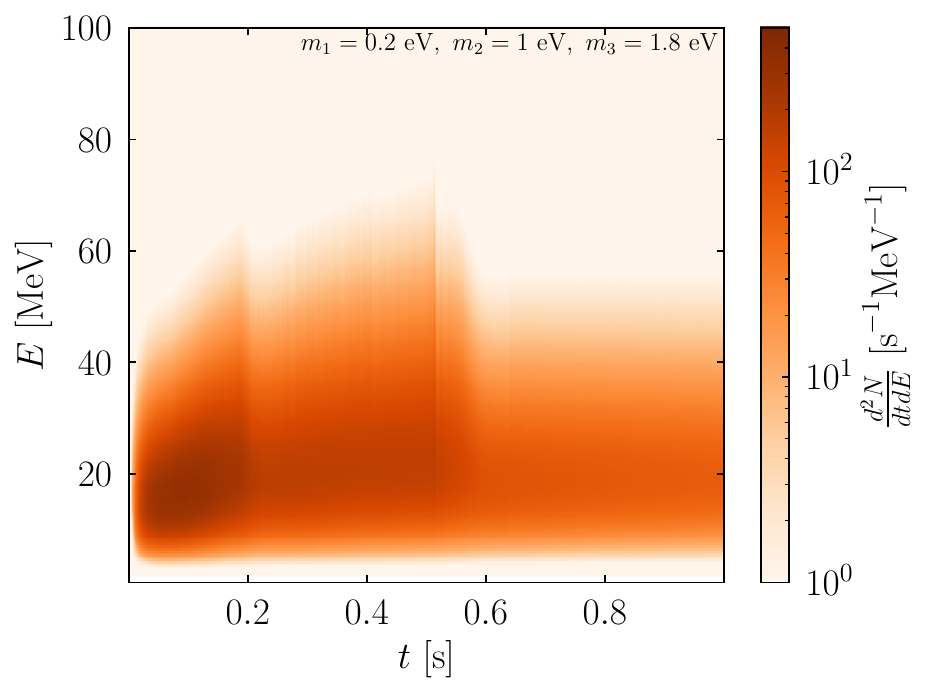}
\includegraphics[width=0.32\textwidth]{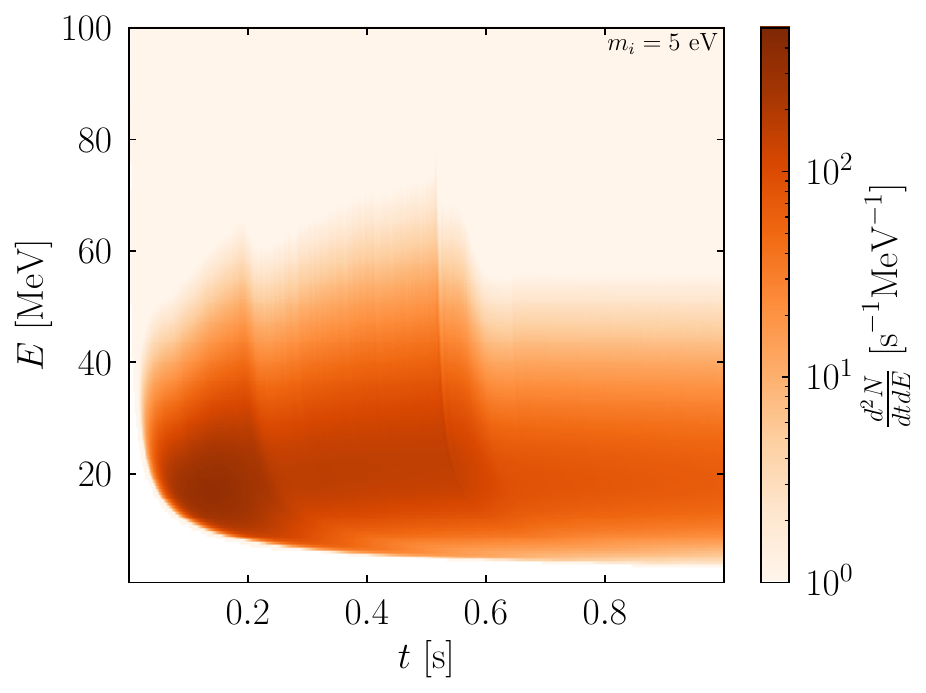}
\includegraphics[width=0.32\textwidth]{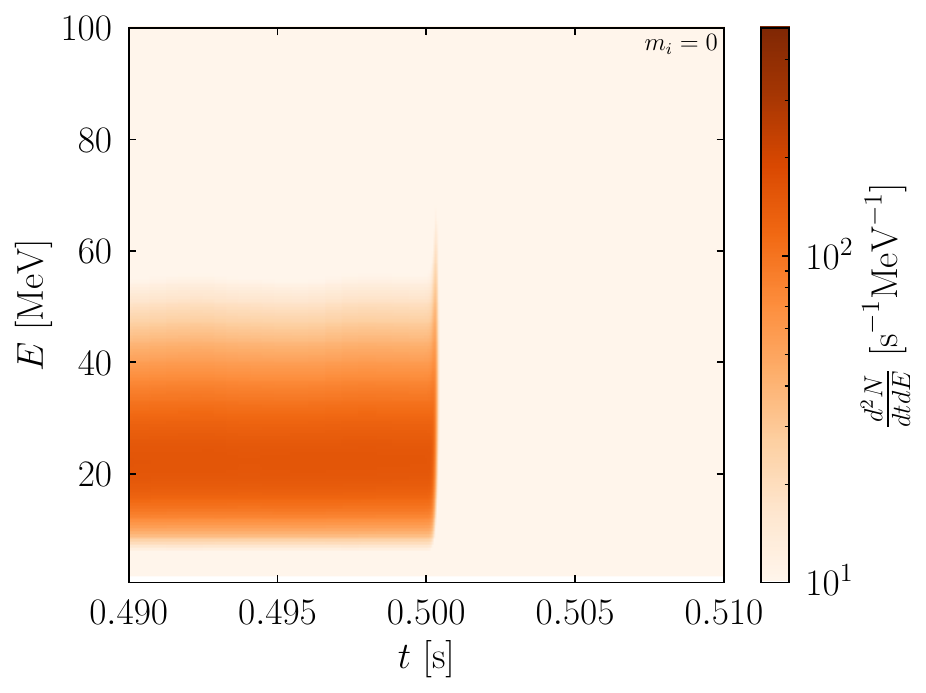}
\includegraphics[width=0.32\textwidth]{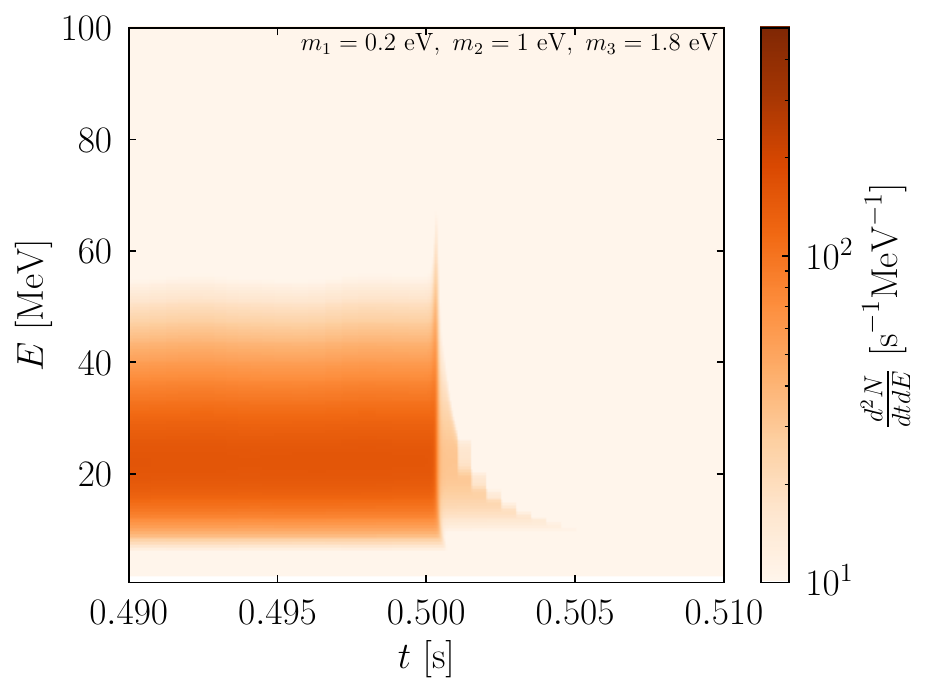}
\includegraphics[width=0.32\textwidth]{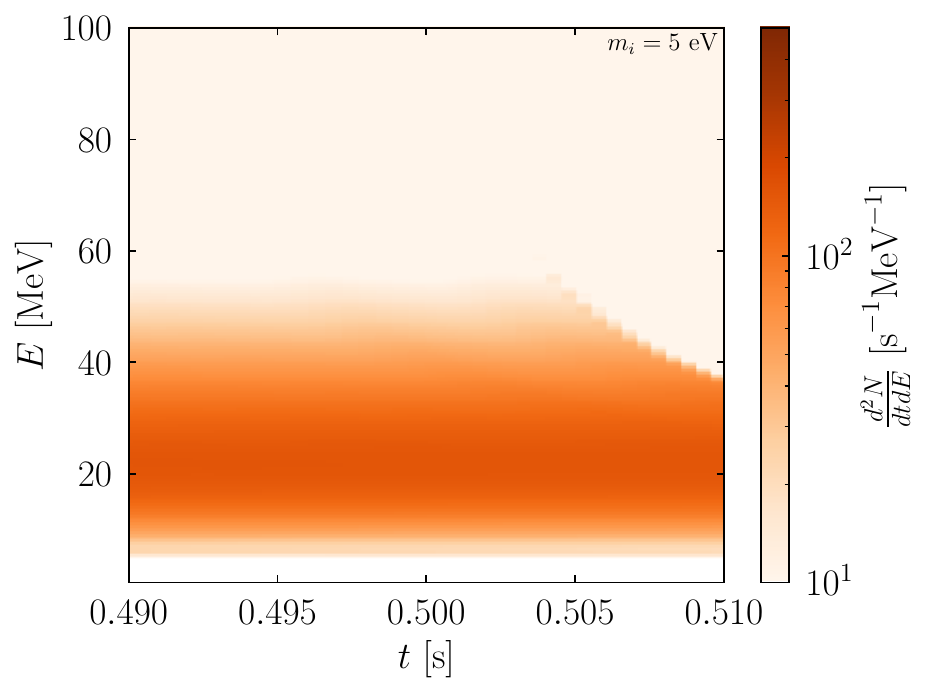}
\caption{The same as fig.~\ref{fig:events QCD} but for the neutronization burst (\textbf{top}) and black hole formation (\textbf{bottom}).
The \textbf{left} panels are for massless neutrinos, the \textbf{middle} panels are for neutrinos with varied high-masses (see subsection \ref{sec:benchmarks} and table \ref{tab:benchmarks}), and the \textbf{right} panels are for very heavy neutrinos.}
\label{fig:events NB BH}
\end{figure*}

In fig.~\ref{fig:events NB BH} we show additional event rate plots for a SN at 10 kpc detected at JUNO in the context of a SN with a neutronization burst only and no QCD phase transition or BH formation (top row) and the case where a BH forms (bottom row).
The three columns represent different mass configurations.
The BH formation case again shows the same feature highlighted in the main text where low energy $\nu_3$'s are visibly delayed due to their larger masses, but the delays of the other states are much more subtle.
The same effects are present in the neutronization burst case, but are somewhat less visible.

\section{Sensitivities in Different Scenarios}
\label{sec:sensitivities appendix}
\begin{figure*}
\centering
\includegraphics[width=0.49\textwidth]{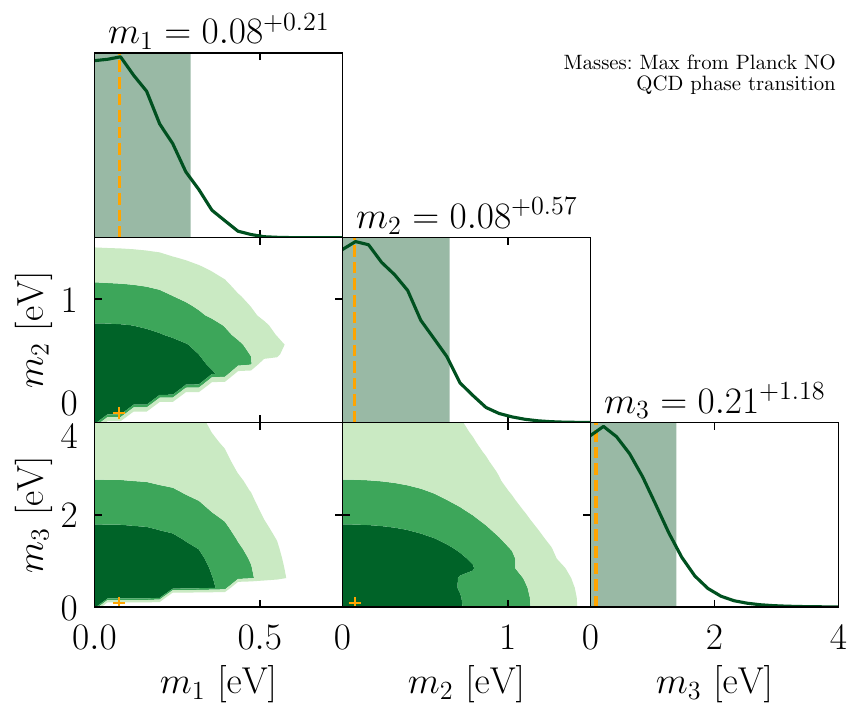}
\includegraphics[width=0.49\textwidth]{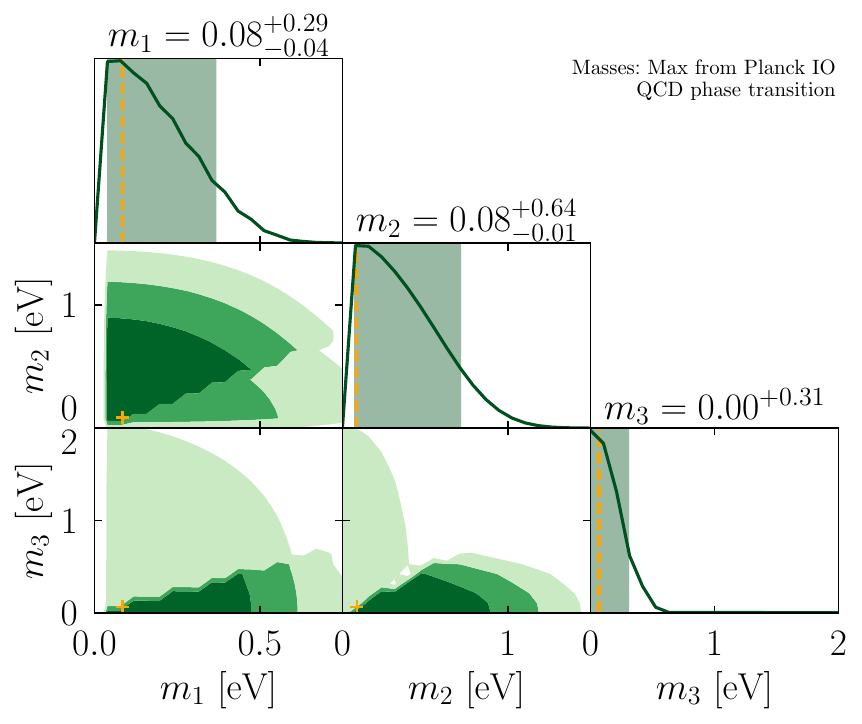}
\caption{The same as fig.~\ref{fig:corner QCD} but for two additional benchmark scenarios: heaviest masses allowed by Planck.}
\label{fig:corner QCD extra}
\end{figure*}

\begin{figure*}
\centering
\includegraphics[width=0.49\textwidth]{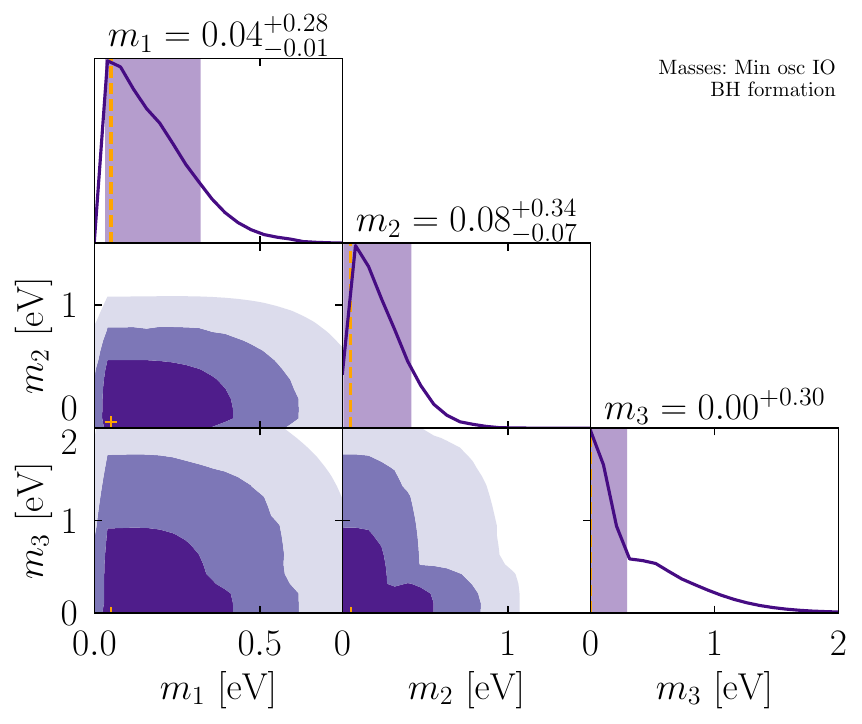}
\includegraphics[width=0.49\textwidth]{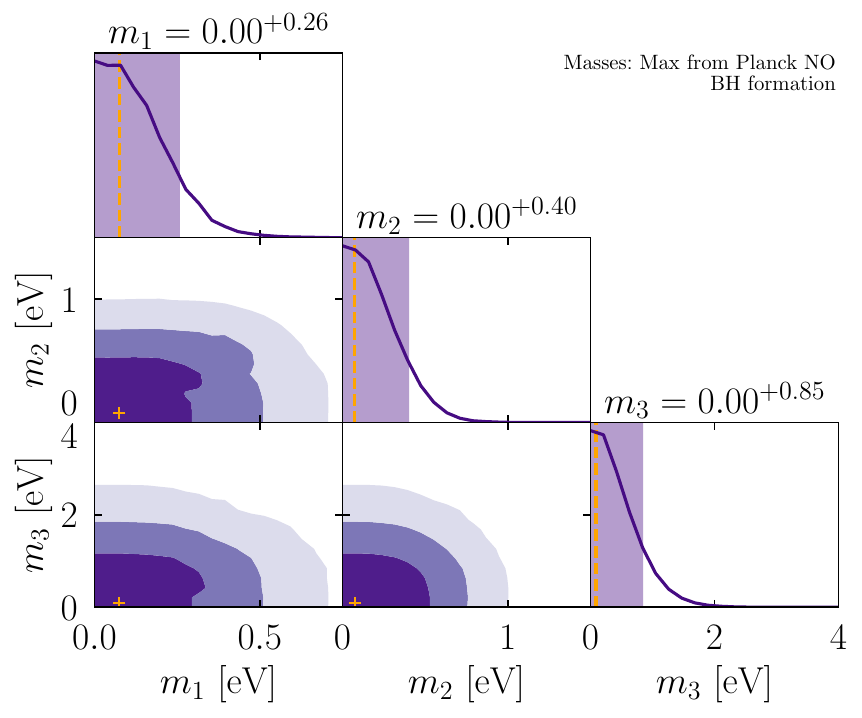}
\includegraphics[width=0.49\textwidth]{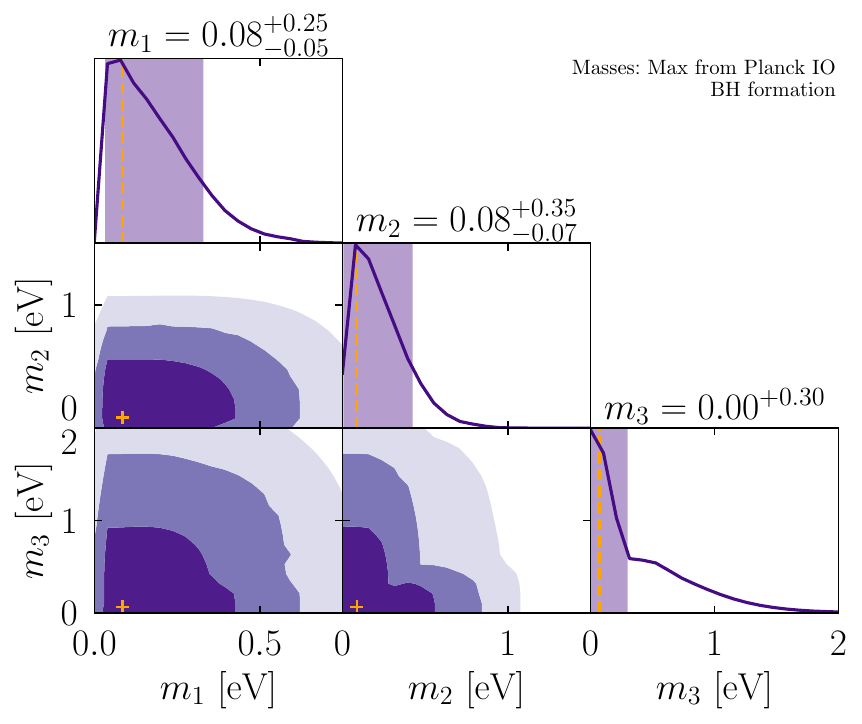}
\includegraphics[width=0.49\textwidth]{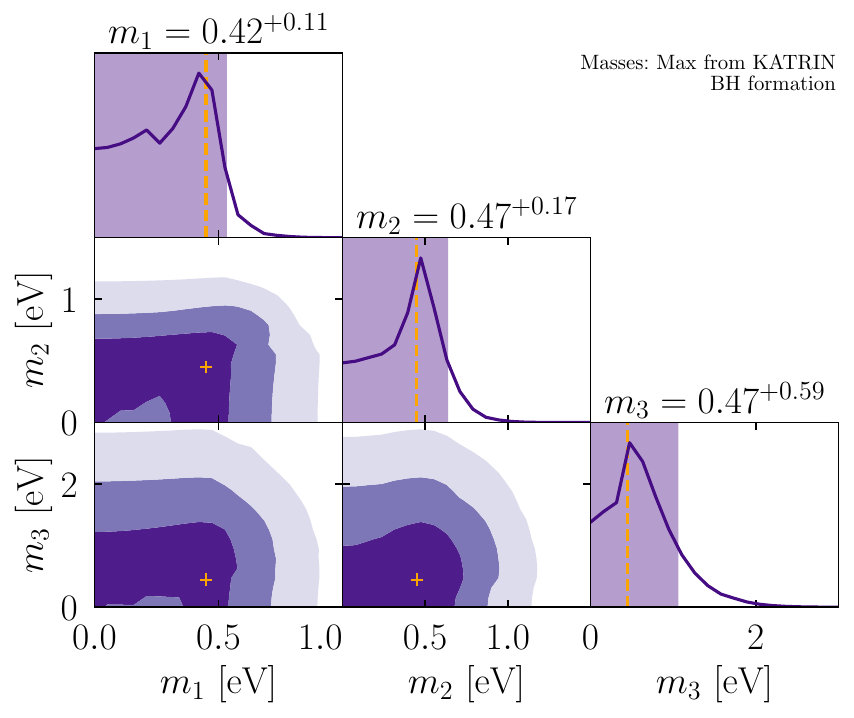}
\caption{The same as fig.~\ref{fig:corner BH} but for four additional benchmark scenarios.}
\label{fig:corner BH extra}
\end{figure*}

\begin{figure*}
\centering
\includegraphics[width=0.49\textwidth]{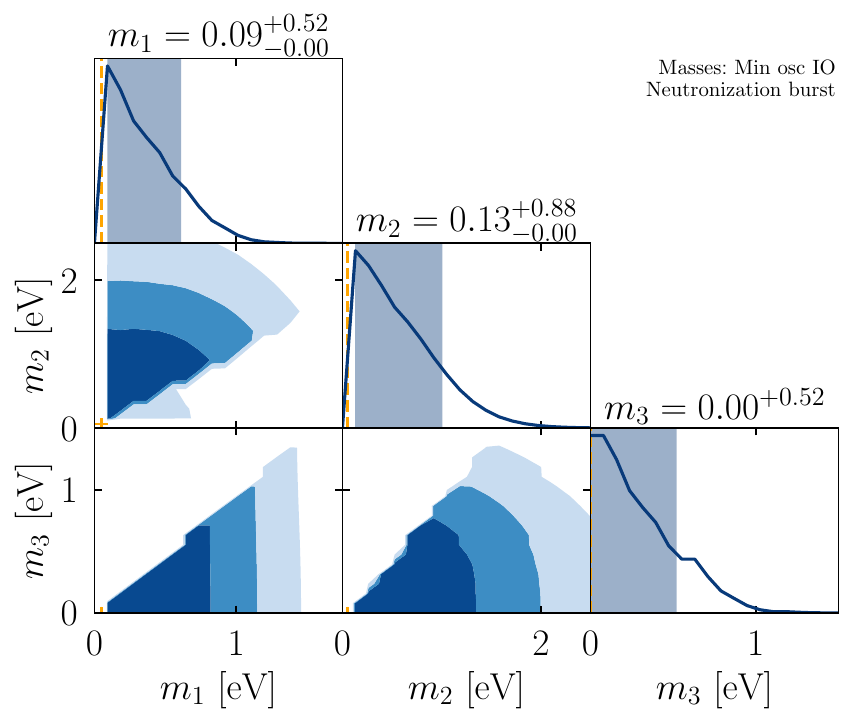}
\includegraphics[width=0.49\textwidth]{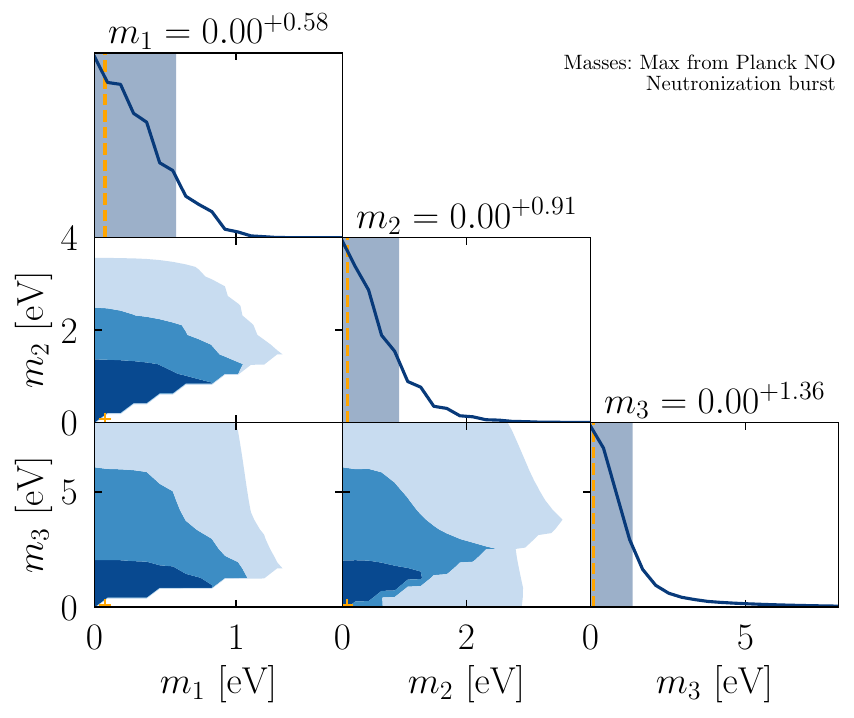}
\includegraphics[width=0.49\textwidth]{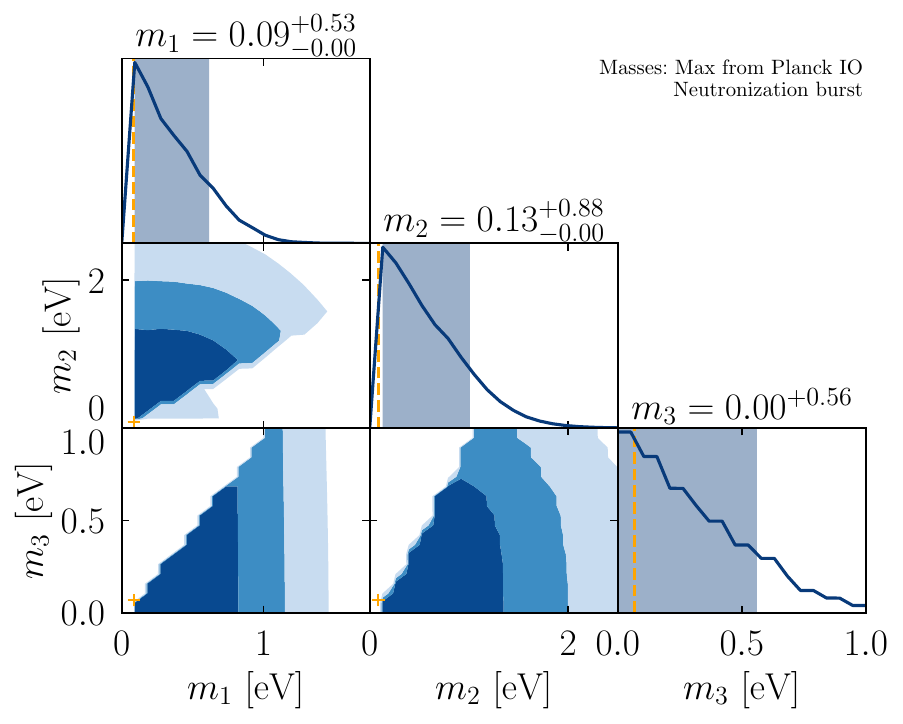}
\includegraphics[width=0.49\textwidth]{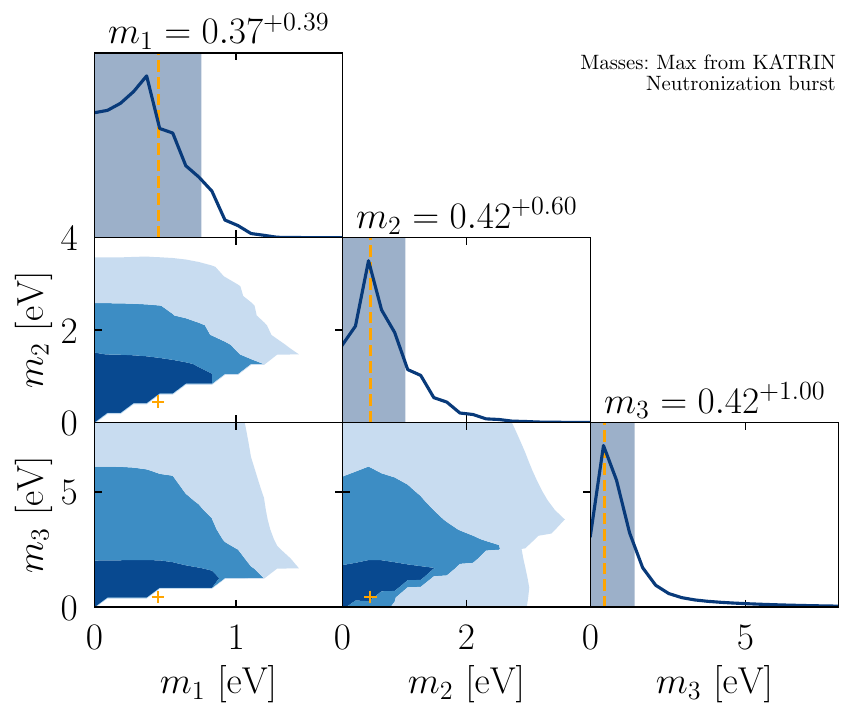}
\caption{The same as fig.~\ref{fig:corner NB} but for four additional benchmark scenarios.}
\label{fig:corner NB extra}
\end{figure*}

In this appendix we show the remaining numerical results for all combinations of benchmarks and physics scenarios within the scenarios for each of the the three mass states in figs.~\ref{fig:corner QCD extra}, \ref{fig:corner BH extra}, and \ref{fig:corner NB extra}.

\bibliography{main}

\end{document}